\documentstyle[12pt]{article}
\newcommand{\nc}{\newcommand}
\nc{\be}{\begin{equation}}
\nc{\ee}{\end{equation}}
\nc{\bea}{\begin{eqnarray}}
\nc{\eea}{\end{eqnarray}}

\nc{\eqn}[1]{{(\ref{#1})}}
\nc{\cA}{{\cal A}}
\nc{\cB}{{\cal B}}
\nc{\cC}{{\cal C}}
\nc{\cD}{{\cal D}}
\nc{\cE}{{\cal E}}
\nc{\cF}{{\cal F}}
\nc{\cG}{{\cal G}}
\nc{\cH}{{\cal H}}
\nc{\cI}{{\cal I}}
\nc{\cJ}{{\cal J}}
\nc{\cK}{{\cal K}}
\nc{\cL}{{\cal L}}
\nc{\cM}{{\cal M}}
\nc{\cN}{{\cal N}}
\nc{\cO}{{\cal O}}
\nc{\cP}{{\cal P}}
\nc{\cQ}{{\cal Q}}
\nc{\cR}{{\cal R}}
\nc{\cS}{{\cal S}}
\nc{\cT}{{\cal T}}
\nc{\cU}{{\cal U}}
\nc{\cV}{{\cal V}}
\nc{\cW}{{\cal W}}
\nc{\cX}{{\cal X}}
\nc{\cY}{{\cal Y}}
\nc{\cZ}{{\cal Z}}
\nc{\simo}[1]{{\stackrel{#1}{\simeq}}}
\nc{\geqo}[1]{{\stackrel{#1}{\geq}}}
\nc{\geo}[1]{{\stackrel{#1}{>}}}
\nc{\guo}[1]{{\stackrel{#1}{\succ}}}

\nc{\rbo}{\raisebox}
\nc{\RR} {\rangle \! \rangle}
\nc{\LL} {\langle \! \langle}
\nc{\rmi}[1]{{\mbox{\small #1}}}
\nc{\eq}{eq.~}
\nc{\nr}[1]{(\ref{#1})}
\nc{\ul}{\underline}
\nc{\mc}{\multicolumn}
\nc{\todo}[1]{\par\noindent{\bf $\rightarrow$ #1}}

\title{\begin{flushright} {\small HD-THEP-95-23} \end{flushright}
\vskip 1cm
Advanced Linked Cluster Expansion.\\
       Scalar Fields at Finite Temperature}
\author{T. Reisz\thanks{Email address reisz@thphys.uni-heidelberg.de}
        \\Institut
        f\"ur Theoretische Physik,\\
	Universit\"at Heidelberg, \\
        Philosophenweg 16, \\
	D-69120 Heidelberg, Germany}

\begin{document}

\maketitle

\begin{abstract}
Linked cluster expansions provide a useful tool for both analytical and
numerical investigations of lattice field theories.
The expansion parameter(s) being the interaction strength(s)
fields at neighboured lattice sites are coupled,
they result into convergent hopping parameter like series
for free energies, correlation functions and in particular susceptibilities.
We consider scalar fields with O(N) symmetric nearest neighbour
interactions on hypercubic lattices with possibly finite extension in
some directions, thus including field theories at
finite temperature $T$.
We improve known and develop new techniques and algorithms to increase the
order $n$ the expansions can be computed to in such a way
that detailed information on critical behaviour can be extracted from
the susceptibility series.
This concerns both simple moments as well as higher correlations
such as 4- and 6-point functions used
to define renormalized coupling constants.
Particular emphasis is done on finite temperature field theory.
In order to be able to measure finite temperature critical behaviour,
the order of explicit computation $n$ has to be sufficiently large
compared to $T^{-1}$ in lattice units.
2- and 4-point susceptibility series are computed up to and including
the 18th order and beyond.
\end{abstract}


\newpage

\section{Introduction}

Most physical problems do not allow for closed analytic solutions.
This is mainly due
to the fact that realistic models can and do evolve considerable
complexity. This is particularly the case with statistical mechanics
and field theory, where the number of degrees of freedom becomes
large or even infinity. Close to criticality those degrees of freedom
interact collectively. One is forced to use approximate methods.

A desirable practice is to construct and apply systematic expansions
about exactly solvable cases such as harmonic oszillators or
zero-dimensional field models. There are essentially two categories.
On the one hand, we use asymptotic methods. Common examples are
saddle point expansions and weak coupling perturbation theory. Those
methods are designated to an accurate quantitative description
of particular sectors, such as the short distance behaviour of
QCD. Their fait is that they are only asymptotically convergent.
On the other hand, if available, convergent expansions have the advantage
that in principle they allow for arbitrary precise measurements within
the domain of convergence.
The price to be paid is that those
expansions typically are more involved, need their own techniques, and
accurate results actually need high orders to be carried out.

We are concerned here with the technique of linked cluster expansion
(LCE) applied
to scalar fields on a hypercubic lattice. It is a special form of the
hopping parameter expansion for the computation of
the free energy, connected correlation functions and susceptibilities.
The expansion is a power series in the strength(s) fields at different
lattice sites are coupled. For standard
actions this just is the nearest neighbour coupling.
The high order behaviour of correlation functions and susceptibilities
allows for a determination of critical couplings and critical exponents.

A good deal to manage the complexity of extreme orders is to use
graph theoretical methods. To every finite order a correlation function
is represented as a sum over a finite number of graphs with
appropriate weights depending on the geometry of the lattice and
the couplings involved. Calculation of extreme orders need techniques
and formalization so that the explicit computation in the end can be done
by a computer. So the main task is the developement of efficient
algorithms and implementations.

The techniques of LCE are of course not new. An elementary introduction
can be found in \cite{Wortis,ID}. The algorithmic formulation of this
expansion for general O(N) symmetric scalar field models
has been pioneered by L\"uscher and Weisz in the famous work
\cite{LW1}, in the course of the solution of lattice $\Phi^4$-theory
in four dimensions \cite{LW2,LW3}.
They succeeded to compute two- and four-point
susceptibility series to 14th order in the hopping parameter
on infinite lattices.

Our main emphasize here is to further develop and improve the
techniques and methods. The aim is to increase the order of explicit
computation for field models both with discrete and continuous
symmetry group in such a way that detailed
knowledge on critical behaviour can be extracted from the LCE series.
A major field of application we have in mind are scalar field theories
at finite temperature.
Those models live on a lattice with toroidal symmetry in temperature
direction with periodic boundary conditions imposed, of extension $L_0$
given by the inverse temperature in lattice units, $L_0=T^{-1}$.
Graphs are able to "feel" temperature only if they can wind around the
torus. There is a least order $n_{\rm min}$ depending on $L_0$, the onset of
$T$-dependence. We expect that $n_{\rm min}$ has to be sufficiently large
compared to $L_0$. That is, critical behaviour at finite $T$
is postponed to higher orders compared to the case of a lattice infinitely
extended in all direction.

The organization of this article is as folllows. We start in Sect.~2
with a reminder to the basics of the LCE of lattice scalar fields.
This concerns motivation, ideas and the problems to be resolved.
LCE series are represented as a sum over (equivalence classes of)
appropriate graphs with associated weights.
Definitions and graph theoretical fundamentals that are used
throughout are introduced.
Sect.~3 is dedicated to the algebraic representation of graphs by
incidence matrices. This is a necessary business to be carried out in
order to generate and compute diagrams numerically.
The main problems to be fixed are uniqueness of the representation, i.e.
1-1 correspondence between equivalence classes of graphs and appropriately
defined canonical incidence matrices, and the construction
of very efficient algorithms to
transform any incidence matrix into its canonical form.
This is at the heart of the problem and necessitates to work around
huge factorials. Introducing extended order relations on vertices
is the way out.
Sect.~4 describes the construction of all the graphs of interest, such as
connected and 1PI ones, along an inheritance tree from so-called
"base classes"
of less complexity and least number of graphs. The order by order
generation is done within the simplest base class only. The computation
of the weights belonging to a graph is discussed.
Sect.~5 summarizes the results.


\section{\label{basic.0}Linked Cluster Expansion}

\subsection{Notations and principles}

We consider a hypercubic $D$-dimensional lattice $\Lambda$, either equal to
$\Lambda_0={\bf Z}^D$ or to $\Lambda_1={\bf Z}/L_0 + {\bf Z}^{D-1}$,
where $D,L_0\in{\bf N}$ and $L_0$ is an even number (finite temperature
lattice with temperature $T=L_0^{-1}$ in lattice units). Periodic
boundary conditions are imposed. The restriction to even $L_0$ is for
convenience and leads to a considerable reduction of the number of
graphs, as will be seen below. To be specific,
the models we discuss are described by the partition function
\be \label{lce.1}
   Z(J,v) = \int \prod_{x\in\Lambda} d^N\Phi(x) \;
   \exp{(-S(\Phi,v)+\sum_{x\in\Lambda}J(x)\cdot\Phi(x) )},
\ee
where $\Phi$ denote a real, $N$-component scalar field, $J$ are
external sources, and
\[
    J(x)\cdot\Phi(x) \; = \; \sum_{a=1}^N J_a(x)\Phi_a(x).
\]
The action is given by
\be \label{lce.2}
   S(\Phi,v) =  \sum_x \stackrel{\circ}{S}(\Phi(x)) + \frac{1}{2}\;
   \sum_{x\not=y\in\Lambda}\sum_{a,b=1}^N \Phi_a(x)v_{ab}(x,y)\Phi_b(y).
\ee
The "ultra-local" (0-dimensional) action $\stackrel{\circ}{S}$ should be
$O(N)$ invariant and respect stability of the partition function
\eqn{lce.1} for sufficiently small $v_{ab}$.
For a $\Phi^4+\Phi^6$ interacting theory,
the standard canonical form is given by
\be \label{lce.3}
   \stackrel{\circ}{S}(\Phi) \; = \; \Phi^2 + \lambda_1 (\Phi^2-1)^2
   + \lambda_2 (\Phi^2-1)^3,
\ee
with $\lambda_2 >0$ or $\lambda_2=0, \lambda_1\geq 0$. Dependence on the
bare ultra-local
coupling constants $\lambda$ is suppressed in the
following. Fields at different lattice sites are coupled by the
hopping term $v(x,y)$. For the case of nearest neighbour interactions,
\be \label{lce.4}
   v_{ab}(x,y) \; = \; \left\{
   \begin{array}{r@{\qquad ,\quad} l }
    2\kappa\;\delta_{a,b}\; & {\rm x,y \; nearest\; neighbour,} \\
    0 & {\rm otherwise.}
   \end{array} \right.
\ee
$\delta$ is the Kronecker symbol. On $\Lambda_1$,
the nearest neighbour property is modulo the torus length.

The generating functional of connected correlation functions
(free energy) is given by
\bea \label{lce.5}
  W(J,v) & = & \ln{ Z(J,v) }, \\
  W_{a_1\ldots a_{2n}}^{(2n)} (x_1,\ldots, x_{2n}) & = &
   <\Phi_{a_1}(x_1) \cdots \Phi_{a_{2n}}(x_{2n}) >^c \nonumber \\
   & = & \left.
   \frac{\partial^{2n}}{\partial J_{a_1}(x_1) \cdots
   \partial J_{a_{2n}}(x_{2n})}
    W(J,v) \right\vert_{J=0},
   \nonumber
\eea
and by a Legendre transform
\bea \label{lce.6}
  \Gamma (M,v) & = & W(J,v) - \sum_{x\in\Lambda} J(x)\cdot M(x) \nonumber \\
   & = & \sum_{n\geq 0} \frac{1}{2n!} \sum_{a_1,\ldots, a_{2n}}
  \Gamma_{a_1\ldots a_{2n}}^{(2n)}(x_1,\ldots, x_{2n}) \;
   M_{a_1}(x_1) \cdots M_{a_{2n}}(x_{2n}), \\
  M_a(x) & = & \frac{\partial W}{\partial J_a(x)}\; ,\quad
   a=1,\ldots ,n \nonumber
\eea
the generating functional of 1-particle irreducible (1PI)
correlations (vertex functional) is defined.
In field theory they are used to define physical coupling constants,
e.g. by
\be \label{lce.7}
  \widetilde\Gamma_{ab}^{(2)}(p,-p) \; = \; - \frac{1}{Z_R} \,
   ( m_R^2 + p^2 + O(p^4) ) \; \delta_{a,b} \quad
   {\rm as} \;p=(p_0=0,\vec{p}\to 0) ,
\ee
where $\;\widetilde{ }\;$ denotes Fourier transform, and
\bea \label{lce.8}
  \widetilde\Gamma_{abcd}^{(4)}(0,0,0,0) & = & - \frac{1}{Z_R^2} \,
   \frac{g_R}{3} C_4(a,b,c,d),\\
  \widetilde\Gamma_{a_1\cdots a_6}^{(6)}(0,\ldots,0) & = &
   - \frac{1}{Z_R^3} \, \frac{h_R}{15} C_6(a_1,\ldots,a_6).
\eea
The $C_n$ denote the O(N)-invariant tensors,
\bea \label{lce.9}
  C_2(a,b) & = & \delta_{a,b} , \\
  C_{2n}(a_1,\ldots , a_{2n}) & = & \sum_{i=2}^{2n} \delta_{a_1,a_i} \,
   C_{2n-2}(a_2,\ldots,\widehat a_i,\ldots,a_{2n}),\quad n\geq 2,
   \nonumber
\eea
($\;\widehat{ }\;$) denotes omission. The more directly accessible
objects from the point of view of the linked cluster
expansion are connected correlation and their susceptibilities.
The latter are defined by
\bea \label{lce.10}
  \delta_{a,b}\; \chi_2 & = & \widetilde W_{ab}^{(2)}(0,0) =
   \sum_x < \Phi_a(x)\Phi_b(0) >^c \\ \label{lce.10.2}
  \delta_{a,b}\; \mu_2 & = & \sum_x g(x) \, < \Phi_a(x)\Phi_b(0) >^c
   = - D_r \, \frac{\partial^2}{\partial p_1^2}
   \left. \widetilde W_{ab}^{(2)}(p,-p) \right\vert_{p=0},
\eea
with $g(x)=\sum_{i=0}^D x_i^2$, $D_r=D$ on $\Lambda_0$ and
$g(x)=\sum_{i=1}^D x_i^2$, $D_r=D-1$ on $\Lambda_1$,
\bea \label{lce.11}
  \frac{1}{3} C_4(a,b,c,d)\; \chi_4 & = &
   \widetilde W_{abcd}{(4)}(0,0,0,0) \\ \label{lce.11.1}
  \frac{1}{15} C_6(a_1,\ldots, a_6)\; \chi_6 & = &
   \widetilde W_{a_1\cdots a_6}^{(6)}(0,\ldots ,0).
\eea
The physical coupling constants as defined
above are related to them by
\bea \label{lce.12}
  m_R^2 & = & 2 D_r \frac{\chi_2}{\mu_2} , \\
  Z_R & = & 2 D_r \frac{\chi_2^2}{\mu_2}, \\
  g_R & = & - \left( \frac{2 D_r}{\mu_2} \right)^2 \; \chi_4 , \\
  h_R & = & - \left( \frac{2 D_r}{\mu_2} \right)^3 \;
   \left( \chi_6 - 10 \frac{\chi_4^2}{\chi_2} \right).
\eea
If the non-ultralocal part of the action is switched off, i.e.~$v=0$,
$S(\Phi,v=0)=\sum_x\stackrel{\circ}{S}(\Phi(x))$, in turn
$W(J,v=0)=\sum_x\stackrel{\circ}{W}(J(x))$. In particular,
\be \label{lce.13}
  W_{a_1\ldots a_{2n}}^{(2n)} (x_1,\ldots, x_{2n}) \; = \; \left\{
   \begin{array}{r@{\qquad ,\quad} l }
    \frac{\stackrel{\circ}{v}_{2n}^c}{(2n-1)!!} C_{2n}(a_1,\ldots , a_{2n})
       & {\rm for \; x_1=x_2=\cdots=x_{2n},} \\
    0 & {\rm otherwise,}
   \end{array} \right.
\ee
with
\be \label{lce.14}
  \stackrel{\circ}{v}_{2n}^c = \left.
   \frac{\partial^{2n}}{\partial J_1^{2n}}\; \stackrel{\circ}{W}(J)
   \right\vert_{J=0}.
\ee
The linked cluster expansion is the Taylor expansion with respect to
$v(x,y)$ about this decoupled case,
\be \label{lce.15}
  W(J,v) = \left. \left( \exp{\sum_{x,y} \sum_{a,b} v_{ab}(x,y)
   \frac{\partial}{\partial \widehat v_{ab}(x,y)}} \right) W(J,\widehat v)
   \right\vert_{\widehat v =0}.
\ee
and the obvious generalizations to connected correlation functions
according to \eqn{lce.5}. Multiple derivatives of $W$ with respect to
$v(x,y)$ are computed by the generating equation \cite{Wortis}
\be \label{lce.15.1}
  \frac{\partial}{\partial v_{ab}(x,y)} W = \frac{1}{2} \left(
   \frac{\partial^2 W}{\partial J_a(x) \partial J_b(y)} +
   \frac{\partial W}{\partial J_a(x)} \frac{\partial W}{\partial J_b(y)}
   \right) .
\ee

The expansion \eqn{lce.15} is convergent under the condition that the
interaction $v(x,y)$ is sufficiently local and weak. In finite
volume this is easily shown, e.g.~by means of the dominated convergence
theorem. In order to get information on collective behaviour, convergence
has to be shown in the infinite volume case. To my knowledge, for this
particular expansion the proof of uniform convergence for large volume
has only recently been done \cite{andreas1}.
In many cases under consideration one restricts attention to couplings of
the form \eqn{lce.4}, i.e~to euclidean symmetric nearest neighbour
interaction of strength $\kappa$ only.
By inspection, for the susceptibilities (\ref{lce.10})-(\ref{lce.11.1})
the coefficients of the series
with respect to $\kappa$ are of equal sign. This remarkable property
implies that the singularity closest to the origin lies on the positive
real $\kappa$-axis. This is identified with the critical point.
In turn it is possible to extract quantitative information on
behaviour even close to criticality from the high order
coefficients of the series \cite{gaunt}.

The evaluation of the Taylor expansion rapidly becomes complex and
cumbersome. A graph theoretical device is the appropriate tool to
keep control of high orders. Correlation functions are represented
as a sum over graphs with appropriate weights. We introduce some basic
concepts.

\subsection{Some Graph Theory}

The notion of a graph used here is a slight specialization of the general
definition \cite{Nakanishi,TR1} adjusted for
the LCE of susceptibilities of scalar field corrrelations.
A graph or diagram is a structure
\be
  \Gamma = (\cL_\Gamma,\cB_\Gamma,E_\Gamma,\Phi_\Gamma),
\ee
where $\cL_\Gamma$ and $\cB_\Gamma\not=\emptyset$ are disjoint sets,
the internal lines and vertices of $\Gamma$, respectively.
$E_\Gamma$ is a map
\bea
  E_\Gamma: \cB_\Gamma & \to & \{0,1,2,\ldots\}, \nonumber \\ \label{lce.16}
   v & \to & E_\Gamma(v),
\eea
that assigns to every vertex $v$ the number of external lines
$E_\Gamma(v)$ attached to it.
The number of external lines of $\Gamma$ is given by
$\sum_{v\in\cB_\Gamma} E_\Gamma(v)$.
Finally, $\Phi_\Gamma$ is the incidence relation that assigns internal lines
to their endpoints. More precisely, let P be the exchange of two vertices,
\bea
  P: \cB_\Gamma\times\cB_\Gamma & \to & \cB_\Gamma\times\cB_\Gamma ,
   \nonumber \\ \label{lce.17}
   (v,w) & \to & (w,v).
\eea
The imbedding map of ordered pairs of vertices onto unordered pairs
is denoted by
$(\bar{\quad})$, $\overline{\cB_\Gamma\times\cB_\Gamma}$
$=(\cB_\Gamma\times\cB_\Gamma)/P$.
We identify $\{(v,v) \vert v\in\cB_\Gamma \}$ with $\cB_\Gamma$ in
the obvious way. Then
\be \label{lce.18}
  \Phi_\Gamma: \cL_\Gamma \to \overline{(\cB_\Gamma\times\cB_\Gamma)}
   \; \setminus \; \cB_\Gamma .
\ee
This maps every internal line onto its (unordered) pairs of endpoints.
Lines are treated as unoriented. We notice the absence of tadpole
lines, i.e.~no line has both of its endpoints equal.
We furthermore set for every $l\in\cL_\Gamma$ with
$\Phi_\Gamma(l)=(v,w)$
\be
  \widehat\Phi_\Gamma (l) = \{ v,w \}.
\ee

For completeness and later use we introduce further notions. External
vertices are those that have external lines attached,
\be \label{lce.19}
  \cB_{\Gamma,ext} =  \{ v\in\cB_\Gamma \; \vert \;
   E_\Gamma(v)\not=0 \},
\ee
whereas internal vertices don't,
$\cB_{\Gamma,int}=\cB_\Gamma\setminus\cB_{\Gamma,ext}$.
For every $v,w\in\cB_\Gamma$,
$m(v,w)$ denotes the number of common lines of $v$ and $w$,
i.e.~the number of elements of
\be \label{lce.20}
  \cM(v,w) = \{ l\in\cL_\Gamma \; \vert \;
   \Phi_\Gamma(l)= \overline{(v,w)} \}.
\ee
The absense of tadpole lines imply that $m(v,v)=0$ for every vertex $v$.
$w$ is a neighbour of $v$ if $m(v,w)\not=0$, and
\be \label{lce.21}
  \cN(v) = \{ w\in\cB_\Gamma \; \vert \; m(v,w)\not=0 \}
\ee
is the set of neighbours of $v$. It has $N(v)$ elements.
Multiple lines do not count here. For every integer $n$,
$n$-vertices are those that have precisely $n$ neighbours,
\be \label{lce.22}
  \cB_\Gamma^{(n)} = \{ v\in\cB_\Gamma \; \vert \;
   N(v)=n \}.
\ee
An $n$-vertex has at least $n$ internal lines attached.

Some topological notions and global properties will be
of major interest in the following.
A path is an ordered non-empty sequence $p=(w_1,\ldots,w_n)$ of
vertices with $m(w_i,w_{i+1})\not=0$ for all $i=1,\ldots ,n-1$. $p$ is
called a path in $\Gamma$ from $w_1$ to $w_n$ of length $n-1$.
If in addition $w_n=w_1$, $p$ is called a loop, with $n-1$ its
number of lines.
If for every pair of vertices $v,w\in\cB_\Gamma$, $v\not=w$,
there is a path from $v$ to $w$, the graph $\Gamma$ is called connected.
If every vertex $v\in\cB_\Gamma$ has an even umber of internal and
external lines attached, i.e.
\be
  l(v) := \sum_{w\in\cB_\Gamma} m(v,w) + E_\Gamma(v) \quad\mbox{even},
\ee
the graph $\Gamma$ is called even. In particular, then,
$\Gamma$ has an even number of external lines.

Two graphs
\be \label{lce.23}
  \Gamma_1 = (\cL_1,\cB_1,E_1,\Phi_1), \;
  \Gamma_2 = (\cL_2,\cB_2,E_2,\Phi_2)
\ee
are called (topologically) equivalent if there are two maps
$\phi_1:\cB_1\to \cB_2$ and $\phi_2:\cL_1\to \cL_2$,
such that
\bea \label{lce.24}
  \Phi_2 \circ \phi_2 &=& \overline\phi_1 \circ \Phi_1, \nonumber \\
  E_2 \circ \phi_1 &=& E_1,
\eea
where $\circ$ means decomposition of maps, and
\bea
  \overline\phi_1: \overline{\cB_1\times\cB_1} & \to &
      \overline{\cB_2\times\cB_2} \nonumber \\ \label{lce.24.1}
   \overline\phi_1(v,w)  & = & (\phi_1(v),\phi_1(w)).
\eea
A symmetry of a graph
$\Gamma = (\cL,\cB,E,\Phi)$ is a pair of maps
$\phi_1:\cB\to\cB$ and $\phi_2:\cL\to \cL$,
such that
\bea \label{lce.25}
  \Phi \circ \phi_2 &=& \overline\phi_1 \circ \Phi, \nonumber \\
  E \circ \phi_1 &=& E.
\eea
The number of those maps is called the symmetry number of $\Gamma$.

The set of equivalence classes of connected graphs with $E$ external
and $L$ internal lines, and with the additional property that every
loop has an even number of lines, is henceforth denoted by
$\cG_E(L)$, and
\be \label{lce.28}
  \cG_E \; := \; \bigcup\limits_{L\geq 0} \; \cG_E(L).
\ee
The constraint on the loop is for later convenience, as will become
clear below.
If in addition the graphs are 1-particle irreducible (1PI),
i.e.~they are still connected
by removing an arbitrary internal line\footnote{We follow standard convention
by using the notion 1PI (1-"particle" irreducibility). It should be kept in
mind, however, that the vertex functional $\Gamma(M,v)$, eqn.~\eqn{lce.6},
is not the generating functional of the LCE diagrams that are 1PI.},
we write $\cG_E^{\rm 1PI}(L)$ and $\cG_E^{\rm 1PI}$.
To any order, the latter graphs are considerably less in number.
Restricting to even graphs only, we define the classes
$\cG_E^{\rm ev}(L)$, $\cG_E^{\rm ev}$ and $\cG_E^{\rm ev,1PI}(L)
= \cG_E^{\rm 1PI}(L)\cap \cG_E^{\rm ev}$, correspondingly.

\subsection{Weights and resummations}

The graphical
representation of the linked cluster expansion of susceptibilities
is an expansion in term of equivalence classes of connected graphs as
defined above. For every such equivalence class we need exactly one
representative.

Every graph $\Gamma$ represents a number, which is called its weight.
If $\Gamma$ is connected and has $E$ external lines, it contributes this
weight to the susceptibility $\chi_E$. For the O(N) scalar field models,
it is computed along the following lines.
Only even graphs need to be considered, otherwise the weight vanishes.

\begin{itemize}
\item All the vertices of $\Gamma$ are placed at lattice sites.
No exclusion principle holds, i.e.~any number of vertices can be placed
at the same site. By assumption, the model lives on the lattice
$\Lambda$, which satisfies translation invariance. An arbitrary selected
vertex is located at a fixed lattice site, avoiding a volume factor.
All the other vertices of the graph are placed arbitrarily.
We remark in passing that it is only at this point where the geometry
of the lattice enters. This includes dimension and topology, such as
toroidal symmetry for the case of a finite temperature lattice.
\item Every internal line of $\Gamma$ is assigned a symmetry label
$a\in\{1,\ldots,N\}$. The line contributes a factor $v_{aa}(x,y)$,
where $x,y\in\Lambda$ are the lattice
sites its endpoints are placed at.
All the external lines get the same fixed symmetry label attached
(usually 1 for convenience).
\item A factor for every vertex $v\in\cB_\Gamma$. Let $2n$ be the sum of
internal and external lines attached to $v$.
Then this factor is given by $\frac{\stackrel{\circ}{v}_{2n}^c}{(2n-1)!!}
C_{2n}(a_1,\ldots , a_{2n})$,
where $a_1,\ldots,a_{2n}$ are the symmetry labels of the internal and
external lines attached to $v$.
\item The sum is taken over all possible assignments of symmetry
labels to lines, and over all possible placements of the vertices at
lattice sites (except for the fixed vertex).
\item Two final factors complete the weight. The first one is given by
$(S(\Gamma))^{-1}$, where $S(\Gamma)$ is the symmetry number of $\Gamma$.
It has the product representation
\be \label{lce.26}
   S(\Gamma) = S_P(\Gamma) \cdot
   \prod_{(v,w)\in\overline{\cB_\Gamma\times\cB_\Gamma}} m(v,w),
\ee
where the integer number $S_P(\Gamma)$ corresponds to the symmetries
of $\Gamma$ under
permutations of vertices. The latter product is due to the exchange of
multiple lines between vertices.
The other final factor counts the various enumerations of the
external lines,
\be \label{lce.27}
    \frac{ E! }{\prod_{v\in\cB_\Gamma} E(v)!}.
\ee
It accounts for the fact that the definition of a graph used here
does not give identities to the external lines
(anonymous factor).
\end{itemize}

For the O(N) models discussed here, summation over all O(N) symmetry
labels and the placement of the vertices on the lattice factorizes,
i.e.~lattice imbedding numbers and O(N) weight factors can be
computed independently for every graph.
Weighted susceptibilities such as the moment $\mu_2$ of (\ref{lce.10.2})
are described by the same diagrams as unweighted ones. The only
difference is that they have the appropriate weight factors inserted
such as $g(x_1,x_2)$, $x_1$ and $x_2$ being the sites of the external
vertices. This, of course, must be done before the sum over the lattice
imbeddings is carried out.

Placing vertices at lattice sites is subject to strong
restrictions by the explicit form of $v(x,y)$. Computations have to take this
into account explicitly.
{}From now on we restrict attention to nearest neighbour interactions
only. That is, $v(x,y)$ is of the form \eqn{lce.4}. The LCE then
is an expansion in powers of $2\kappa$. The order of $\kappa$ a graph
contributes to is given by its number of lines. Furthermore,
the final imbedding of the graphs is done onto the hypercubic lattice
$\Lambda$. Vertices connected by at least one line have to be placed
at nearest neighbour lattice sites. A necessary condition for a
non-vanishing imbedding to
exist is that every loop of this graph has an even number
of lines\footnote{We remember that $L_0$, the possibly finite extent
of the lattice in one direction, is always chosen to be even.}.
It is meaningful to take this into account from the beginning.
This implies a considerable reduction on the number of diagrams
to be considered, as we have done with the definition of $\cG_E$
above.

Connected graphs and in turn susceptibilities can be explicitly written
in terms of 1PI ones.
\bea \label{lce.28.10}
  \chi_2 & = & \frac{\chi_2^{\rm 1PI}}{1-(2\kappa)(2D)\chi_2^{\rm 1PI}} , \\
   \label{lce.28.11}
  \mu_2 & = & \frac{\mu_2^{\rm 1PI}+(2\kappa)(2D_r)(\chi_2^{\rm 1PI})^2}
     {(1-(2\kappa)(2D)\chi_2^{\rm 1PI})^2} , \\
  \chi_4 & = & \frac{\chi_4^{\rm 1PI}}
     {(1-(2\kappa)(2D)\chi_2^{\rm 1PI})^4} , \\ \label{lce.29}
  \chi_6 & = &  \frac{1}{(1-(2\kappa)(2D)\chi_2^{\rm 1PI})^6}\;
    \left( \chi_6^{\rm 1PI} + 10 (2\kappa)(2D)
     \frac{(\chi_4^{\rm 1PI})^2}{ 1-(2\kappa)(2D)\chi_2^{\rm 1PI}}
    \right).
\eea
The 1PI susceptibilities admit the series representation
\bea \label{lce.30}
  \chi_E^{\rm 1PI} & = & \sum_{L\geq 0} a_{E,L}^{\rm 1PI}(\lambda)
   (2\kappa)^L , \\
  a_{E,L}^{\rm 1PI}(\lambda,\sigma) & = & \sum_{G\in\cG_E^{\rm ev,1PI}(L)}
   a_E^{\rm 1PI}(\lambda,G),
\eea
where $a_E^{\rm 1PI}(\lambda,G)$ is the weight of the diagram
$G$ that is computed as described above, except for the factor
$(2\kappa)^L$ that we have written out explicitly.
Similar representations hold with the obvious modifications for
e.g.~$\mu_2^{\rm 1PI}$.
According to the identities (\ref{lce.28.10}f) it is sufficient
to restrict attention to 1PI graphs only.
Actually, this simplification implies even more than just a reorganization
of diagrams. It turns out that critical surfaces are computed
from the series expansion of $\chi_2^{\rm 1PI}$ with much better
accuracy than from the series of $\chi_2$.  They are determined by
the roots of the denominator of  \eqn{lce.28.10}.

Further simplification arises by classifying diagrams in terms of even
more stringent properties \cite{Wortis}. An essential characterization
that we adhere to is 1-vertex irreducibility (1VI).
A graph $\Gamma$ is called 1VI if the following condition is satisfied.
Remove any vertex $v$ of $\Gamma$, together with the external and
internal lines attached to $v$. We denote the resulting graph by
$\Gamma_v = (\cL_v,\cB_v,E_v,\Phi_v)$.
Then, for every $w\in\cB_v$ there exists an external vertex $u\in\cB_v$
and a path from $u$ to $w$ in $\Gamma_v$.
In other words, every connected component of $\Gamma_v$ has at least
one external line left attached to one of its vertices.
We write
\be \label{lce.31}
  \cS_k(L) = \{ \Gamma\in\cG_k^{\rm 1PI}(L) \;\vert\;
   \Gamma\; \mbox{is 1VI}\; \}
\ee
and $\cS_k^{\rm ev}(L)=\cS_k(L)\cap\cG_k^{\rm ev}$,
for the set of graphs that are both 1PI and 1VI.
On the other hand, renormalized moment diagrams are 1PI graphs that have
exactly one external vertex,
\be \label{lce.32}
  \cQ_k(L) = \{ \Gamma\in\cG_k^{\rm 1PI}(L) \;\vert\;
   \mbox{there is $v\in\cB_\Gamma$ with}\; E_\Gamma(v)=k\; \},
\ee
and $\cQ_k^{\rm ev}(L)=\cQ_k(L)\cap\cG_k^{\rm ev}$.
With both of these notions, susceptibilities are represented as
\be \label{lce.33}
  \chi_E^{\rm 1PI} = \sum_{L\geq 0} \; \sum_{G\in\cS_E^{\rm ev}(L)}
   a_E^{\rm 1VI}(\lambda,\kappa,G) (2\kappa)^L ,
\ee
where the $a_E^{\rm 1VI}$ are computed in the same way as are the
$a_E^{\rm 1PI}$ above with the following exception only.
The vertex couplings
$\stackrel{\circ}{v}_{2n}^c$ are replaced by the
renormalized moments
\be
  \stackrel{\circ}{v}_{2n}^c \to v_{2n}^c =
   \sum_{L\geq 0} \; \sum_{G\in\cQ_{2n}^{\rm ev}(L)}
   a_{2n}^{\rm 1PI}(\lambda,G) (2\kappa)^L ,
\ee
and the $a^{\rm 1PI}$ are computed as in \eqn{lce.30} before.

1VI and renormalized moments provide a computational device, which we
can think of as a kind of product representation of 1PI graphs by
1VI diagrams and renormalized moments.
In the final end, after having determined all the weights of the graphs,
the coefficients are reorganized to give the coefficients
$a_{E,L}^{\rm 1PI}(\lambda)$.


\section{Algebraic representation of graphs.}

\subsection{Incidence Matrices}

In order to investigate graphs algorithmically and by a computer we need a
convenient algebraic representation. Examples are linked list connectivity
structures or incidence matrices \cite{mehlhorn,sedgewick}.
The latter appear to be more useful for numerical operations.

Graphs of the LCE do not have tadpole lines. This allows for the
following definition.
Let $\Gamma$ be a graph, $V_\Gamma$ the number of its vertices.
Then an incidence matrix $I_\Gamma$ of $\Gamma$ is defined as follows.
\begin{itemize}
\item Enumerate the vertices of $\Gamma$, so that
\be \label{alg.1}
  \cB_\Gamma = \left\{ v_i \;\vert\; i=1,\ldots,V_\Gamma \right\}.
\ee
In particular, $v_i\not= v_j$ for $i\not= j$.
\item For $i,j=1,\ldots ,V_\Gamma$, $i\not= j$, define
\be \label{alg.2}
  I_\Gamma (i,j) = m(v_i,v_j),
\ee
where $m(v_i,v_j)$ is the number of lines between $v_i$ and $v_j$.
\item The diagonal elements are given by the number of external lines
attached to the vertices,
\be\label{alg.3}
  I_\Gamma(i,i) = E(v_i)\; ,\quad
   i=1,\ldots,V_\Gamma.
\ee
\end{itemize}

The incidence matrix is a symmetric $V_\Gamma\times V_\Gamma$ matrix,
$I_\Gamma\in M(V_\Gamma\times V_\Gamma)$. Most
algorithms need to deal with e.g.~the right upper triangle only.

The construction above does not provide a unique definition of
an incidence matrix of a graph. Any other
enumeration of its vertices correspond to a non-trivial permutation
$\pi\in\Pi_{V_\Gamma}$ and leads to the incidence matrix
\be \label{alg.4}
  I_\Gamma^\pi (i,j) = I_\Gamma(\pi(i),\pi(j)),
\ee
i.e.~$I_\Gamma$ is unique only modulo simultaneous permutations of its
rows and columns. Those permutations that leave $I_\Gamma$
unchanged constitute a finite subgroup of $\Pi_{V_\Gamma}$. The number
of elements of this group is equal to $S_P(\Gamma)$.
Two graphs $\Gamma_1$ and $\Gamma_2$ are topologically equivalent if and
only if they have the same number of vertices $V$, say, and
their respective incidence matrices (with respect to an arbitrary
initial labelling of the vertices) satisfy
\be \label{alg.5}
  I_{\Gamma_1}^\pi = I_{\Gamma_2},
\ee
for some $\pi\in\Pi_V$.

During the process of graph
generation discussed in the next section, many equivalent graphs will be
constructed, but for every equivalence class we have to keep only one
graph. To decide whether two graph are equivalent according to
\eqn{alg.5} will need factorials of permutations to be carried out.
This is too expensive for high order graphs. What is needed is a unique
representation of graphs by incidence matrices. Furthermore, the
computation of those incidence matrices should be as efficient as
possible. Definitely huge factorials have to be avoided.

The representation of graphs by incidence matrices can be made unique
if a total order relation on $M(V\times V)$ is chosen for every $V$,
and in turn the maximum over all simultaneous permutations of rows
and columns is taken. Then, for an arbitrary initial labelling,
we get a canonical representation by
\be \label{alg.6}
  \widehat I_\Gamma := \max_{\pi\in\Pi_{V_\Gamma}} I_\Gamma^\pi,
\ee
and the maximum is with respect to the total order relation. A convenient
choice is as follows. Let $A,B\in M(V\times V)$. Then
\be \label{alg.7}
   A > B  \stackrel{\rm DEF}{\Longleftrightarrow}
   \begin{array}{ c }
   \mbox{there is}\; i,j\in\{1,\ldots,V\}\;\mbox{with}\;
   A(i,j)>B(i,j), \;\mbox{and}\nonumber \\
   A(k,l)=B(k,l)\;\mbox{for all}\;k=i,l<j\;
    \mbox{and for all}\; k<i.
   \end{array}
\ee
Still there are huge factorials involved to construct the canonical form
of incidence matrices. The breaktie solution here is to introduce an order
relations for the vertices of a graph first. Topological properties
of the graph are taken into accont here, local and less local ones,
as we will discuss below. Ordering
cannot be complete, at least those vertices need to stay
unordered relative to each other whose exchange correspond to a
symmetry of the graph. Once the order relation has been chosen, a canonical
incidence matrix can be defined by subsequently allowing
in \eqn{alg.6} only those vertices to be exchanged that stay unordered
relative to each other.

For any graph $\Gamma$ let $\cO$ be an order relation on its set of
vertices $\cB_\Gamma$. For any pair of vertices $v,w\in\cB_\Gamma$,
$v\not= w$, either $v\geo{\cO}w$, $w\geo{\cO}v$, or $v\simo{\cO}w$
if neither $v\geo{\cO}w$ nor $w\geo{\cO}v$.
The enumeration of the vertices then is any one such that
$v_1\guo{\cO} v_2 \guo{\cO} v_3 \guo{\cO} \cdots
\guo{\cO} v_{V_\Gamma}$, where $\guo{\cO}$ means
$\geo{\cO}$ or $\simo{\cO}$, leading to some $I_\Gamma$.
The canonical incidence matrix of $\Gamma$ then is defined by
\be \label{alg.8}
  \widehat I_\Gamma^\cO := \max_{\pi\in\Pi_{V_\Gamma}^\cO} I_\Gamma^\pi,
\ee
with $\Pi_{V_\Gamma}^\cO$ the subset of $\Pi_{V_\Gamma}$ whose elements
correspond to the exchange of those vertices only that stay unordered
relative to each other with respect to $\cO$. Furthermore, the number
of permutations $\pi$ with $I_\Gamma^\pi=\widehat I_\Gamma^\cO$
is equal to the (partial) symmetry factor $S_P(\Gamma)$.

This definition of an incidence matrix satisfies all the above criteria.
The particular form depends on the order relation $\cO$. The expense of
ordering vertices is with some power of the number of vertices, much
cheaper than factorials.

The heart of the problem is to find an appropriate order relation $\cO$
on the vertices of a graph. The number of remaining permutations of
\eqn{alg.8} should be as small as possible. Typically it is still some
factorial, but smaller than the original one. To some order of the LCE
we succeed with a chosen vertex ordering, i.e~all the graphs of
a particular correlation function can be generated with acceptable
costs. With increasing number of lines, however, the problem of large
factorials reappears, and one is forced to extend the order relation.
The cardinality of $\Pi_{V_\Gamma}^\cO$ is bounded from below by the
partial symmetry number $S_P(\Gamma)$. There is much space left towards
this ideal case.

For later convenience we introduce the following definition.
Let $\cO_i$, $i=1,\ldots,n$, be order relations on the set of vertices
$\cB_\Gamma$ of a graph $\Gamma$. Then we define an order relation
$\cO = <\cO_1,\ldots,\cO_n>$ on $\cB_\Gamma$ as follows.
For $v,w\in\cB_\Gamma$, $v$ is called larger than $w$ with respect to
$\cO$, $v\geo{\cO}w$, if there is $i\in\{1,\ldots,n\}$ such that
\be \label{alg.9}
   v\geo{\cO_i}w
\ee
and for all $j<i$
\be \label{alg.10}
   v\simo{\cO_j}w.
\ee
If for all $i\in\{1,\ldots,n\}$ $v\simo{\cO_j}w$,
$v$ and $w$ stay unordered with respect to $\cO$, $v\simo{\cO}w$.
We notice that the order relation $\cO$ depends on the order of the
sequence $\cO_1,\ldots,\cO_n$.

In connection with the LCE, the first order relation on the vertices is
due to L\"uscher and Weisz \cite{LW1}. It takes into account all local
properties of the graph in an optimal way. That is,
the order number assigned to a vertex $v$ is based on the connections
to its neighbours $\cN(v)$. Let
$\nu_i(v)$ denote the number of vertices of a graph $\Gamma$ that have
precisely $i$ lines in common with the vertex $v$. Then
\be \label{alg.11.1}
  \cO_{LW} = <\cO_\alpha ,\cO_\beta>,
\ee
where for all $v,w\in\cB_\Gamma$,
\bea \label{alg.11}
  v \geo{\cO_\alpha}  w  & \Longleftrightarrow &
   E(v) > E(w), \\
  v \geo{\cO_\beta}  w  & \Longleftrightarrow &
   \mbox{there is}\; i\geq 1 \;\mbox{with}\;\nu_i(v)>\nu_i(w)
   \;\mbox{and}\nonumber \\
   & & \nu_j(v) = \nu_j(w) \;\mbox{for all}\;j>i .
\eea
By means of this order relation the graphs of the
two- and four-point correlations can be generated up to and including
the 14th order graphs. This order appears as some natural limit, due to
the kind of slowing down mentioned above.
As we have discussed in Sect.~1, in order to obtain most accurate
quantitative information on critical behaviour from the LCE
series, we want to compute them up to the 18th order.
In particular we expect to be forced to do so on finite temperature lattices.

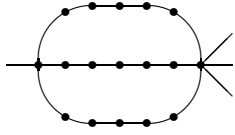
\begin{figure}[htb]
\caption{\label{preord} Motivation for extended vertex ordering. The graph
$\Gamma$ shown has many 2-vertices. They stay unordered relative to
each other by the order relation $\cO_{LW}$.}

\begin{center}
\setlength{\unitlength}{0.6cm}
\begin{picture}(4.0,5.0)

\put(2.9000,1.8000){\oval(3.600,2.600)}
\put(1.1000,1.8000){\circle*{0.16}}
\put(4.7000,1.8000){\circle*{0.16}}

\put(1.7000,2.9700){\circle*{0.16}}
\put(2.3000,3.1000){\circle*{0.16}}
\put(2.9000,3.1000){\circle*{0.16}}
\put(3.5000,3.1000){\circle*{0.16}}
\put(4.1000,2.9700){\circle*{0.16}}

\put(1.7000,0.6300){\circle*{0.16}}
\put(2.3000,0.5000){\circle*{0.16}}
\put(2.9000,0.5000){\circle*{0.16}}
\put(3.5000,0.5000){\circle*{0.16}}
\put(4.1000,0.6300){\circle*{0.16}}
\put(1.1000,1.8000){\line(1,0){3.6000}}
\put(1.7000,1.8000){\circle*{0.16}}
\put(2.3000,1.8000){\circle*{0.16}}
\put(2.9000,1.8000){\circle*{0.16}}
\put(3.5000,1.8000){\circle*{0.16}}
\put(4.1000,1.8000){\circle*{0.16}}
\put(1.1000,1.8000){\line(-1,0){0.7}}
\put(4.7000,1.8000){\line(1,0){0.7}}
\put(4.7000,1.8000){\line(1,1){0.7}}
\put(4.7000,1.8000){\line(1,-1){0.7}}

\end{picture}
\end{center}

\end{figure}

The problem that arises becomes clear already by considering
the 18th order graph
$\Gamma$ of Fig.\ref{preord}. The original 17!~number of permutations
to be carried out in order to construct the incidence matrix according to
\eqn{alg.6} is reduced after $\cO_{LW}$ to 15!, which still is a large
number. This is
because the graph has a large number of 2-vertices. On the other hand,
there are obvious subsets of 2-vertices whose permutations cannot be
part of a symmetry. Those vertices should be further distinguished by
the order relation on $\cB_\Gamma$. For instance, as an immediate
criterion we can
use the distance of such a vertex along a chain of 2-vertices towards
the next non-2-vertex.

Before we discuss the extension of the order relation $\cO_{LW}$,
we notice that for every pair of vertices $v,w\in\cB_\Gamma$ with
$v\in\cB_{\Gamma,int}^{(2)}$, $w\not\in\cB_{\Gamma,int}^{(2)}$,
we have either $v\geo{\cO_{LW}}w$ or $w\geo{\cO_{LW}}v$.
That is, $\cO_{LW}$ separates internal 2-vertices from those that are
not. Extension of $\cO_{LW}$ can thus be done for internal 2-vertices
and their complement separately.

Henceforth we need the following definition.
Let $\cV_1,\cV_2\subseteq\cB_\Gamma$ be complementary subsets,
$\cV_1\cap\cV_2=\emptyset$, $\cV_1\cup\cV_2=\cB_\Gamma$,
and let $\cO_1$ and $\cO_2$ be order relations defined on
$\cV_1$ and $\cV_2$, respectively.
Furthermore, let $\cO_0$ be an order relation on $\cB_\Gamma$ such that
for all $v_1\in\cV_1$, $v_2\in\cV_2$ either
$v_1\geo{\cO_0}v_2$ or $v_2\geo{\cO_0}v_1$. Then we define an order
relation $\cO = < \cO_0, [\cO_1,\cO_2] >$ as follows.
For all $v,w\in\cB_\Gamma$,
\be \label{alg.12}
   v \geo{\cO} w \; \Longleftrightarrow \; \left\{
   \begin{array}{r@{\qquad ,\quad} l }
    v\geo{\cO_0}w & \mbox{if $v\in\cV_1$, $w\in\cV_2$ or vice versa,} \\
    v\geo{\cO_1^\prime}w & \mbox{if $v,w\in\cV_1$,} \\
    v\geo{\cO_2^\prime}w & \mbox{if $v,w\in\cV_2$,}
   \end{array} \right.
\ee
where $\cO_1^\prime = < \cO_0, \cO_1 >$,
$\cO_2^\prime = < \cO_0, \cO_2 >$.
If neither $v\geo{\cO}w$ nor $w\geo{\cO}v$, we write $v\simo{\cO}w$.
This can happen only if $v,w$ both belong to $\cV_1$ or both to $\cV_2$.

\subsection{Extended vertex ordering for non-ring graphs.}

In the following we restrict attention to connected graphs only.
We have to distinguish between ring graphs and their complement,
non-ring graphs. By definition, a connected graph $\Gamma$
is a ring graph if $\Gamma$ has at least
three vertices, and $\cB_\Gamma \not= \cB_{\Gamma,int}^{(2)}$. That is,
$\Gamma$ has at least one vertex that is not an internal 2-vertex.
We discuss extended vertex ordering of non-ring graphs first, the
modifications for ring graphs are described afterwards.
Let us suppose in this section that $\Gamma$ is not a ring graph.

We define the order relation $\cO$ used with \eqn{alg.8} to define
the canonical incidence matrix by
\be \label{alg.13}
   \cO = < \cO_{LW}, [\cO_{ni},\cO_{int}] >,
\ee
where $\cO_{int}$ and $\cO_{ni}$ are convenient order relations
on the set of internal 2-vertices and its complement, respectively.
As we shall see they can be chosen in such a way that the vertex
ordering is considerably enhanced. We are now going to define
$\cO_{ni}$ and $\cO_{int}$.

A path $p=(w_1,w_2,\ldots,w_n)$ is called a 2-chain of $w_1$ if
$w_2,\ldots,w_{n-1}\in\cB_{\Gamma,int}^{(2)}$ are mutually
distinct. If the last vertex is
not an internal 2-vertex, $w_n\not\in\cB_{\Gamma,int}^{(2)}$, $p$ is
called maximal 2-chain of $w_1$.
In this case we assign a tripel of two integers and a vertex to p by
\bea
  \mu(p) & = & n-1 , \\ \label{alg.14}
  \lambda(p) & = & \sum_{i=1}^{n-1} m(w_i,w_{i+1}) ,\;\mbox{and} \\
  e(p) & = & w_n.
\eea
For any $v\in\cB_{\Gamma,int}^{(2)}$ there are precisely two distinct
maximal 2-chains. Let us denote them by $c_1(v)$ and $c_2(v)$.
They intersect in $v$ and possibly the last vertex only.
For later use we set
\be \label{alg.15}
  \mu(v) = \max_{i=1,2} \; \mu(c_i).
\ee
Let $\cO_{non}$ be an order relation on
$\cB_\Gamma\setminus\cB_{\Gamma,int}^{(2)}$. It induces an order on the
maximal 2-chains $\{c_1(v),c_2(v)\}$ as follows.
For $i,j\in\{1,2\}$, $i\not=j$,
\be \label{alg.16}
  c_i(v) > c_j(v) \; \Longleftrightarrow \;
   \begin{array}{ c }
    \mu(c_i) < \mu(c_j) \; \mbox{, or} \\
    \mu(c_i) = \mu(c_j) \; \mbox{and} \; \lambda(c_i)>\lambda(c_j)
     \; \mbox{, or} \\
    \mu(c_i) = \mu(c_j) \; \mbox{and} \; \lambda(c_i)=\lambda(c_j)
     \; \mbox{and}\; e(c_i)\geo{\cO_{non}}e(c_j) .
   \end{array}
\ee
If unique, we choose "short and long chains" by
\bea \label{alg.17}
  c_s(v) & = & \max_{i=1,2} \; c_i(v), \\
  c_l(v) & = & \min_{i=1,2} \; c_i(v).
\eea
Roughly speaking, $c_s(v)$ gives a stronger binding of $v$ to
$e(c_s)$ than $c_l(v)$ to $e(c_l)$ (shorter chain, more lines).
If neither $c_1(v)>c_2(v)$ nor $c_2(v)>c_1(v)$, we choose any one of the
maximal 2-chains as $c_s(v)$ and the other one as $c_l(v)$.
This freedom is allowed because all that will be used is the order
relation \eqn{alg.16}. We remark in passing that $e(c_1)$ and
$e(c_2)$ can be identical.

Finally, $\cO_{non}$ induces through the definition \eqn{alg.16}
an order relation
$\cO_{int}$ on $\cB_{\Gamma,int}^{(2)}$ by
$\cO_{int} = < \cO_{i1}, \cO_{i2} >$, with
\bea \label{alg.18}
  v\geo{\cO_{i1}}w & \Longleftrightarrow &
   c_s(v) > c_s(w) , \\
  v\geo{\cO_{i2}}w & \Longleftrightarrow &
   c_l(v) > c_l(w) ,
\eea
for all internal 2-vertices $v,w$.
This is defined once $\cO_{non}$ is fixed. One possibility is to define
$\cO_{non}$ as equal to $\cO_{LW}$. But it appears that only part of
the relations lead to considerable enhancement of ordering, other ones
in very exceptional cases only.
Furthermore, in addition to supplementing the ordering of internal
2-vertices, we also extend the order relations on the other vertices.
Possible criteria are properties of attached 2-chains.
Let us call such a relation $\cO_{ni}$.
Then we let
$\cO_{non} = < \cO_\alpha,\cO_\gamma,\cO_\delta, \cO_{ni} >$,
where $\cO_\alpha$ is defined by \eqn{alg.11}, and
\bea
  v\geo{\cO_\gamma}w & \Longleftrightarrow &
   N(v)>N(w) , \\
  v\geo{\cO_\delta}w & \Longleftrightarrow &
   \max_{\nu_k(v)\not=0} k  > \max_{\nu_k(w)\not=0} k,
\eea
for all $v,w\in\cB_\Gamma\setminus\cB_{\Gamma,int}^{(2)}$.

All that is left to be done is to define $\cO_{ni}$.
In the first place $\cO_{ni}$ supplements $\cO_{LW}$ for vertices that
are not internal 2-vertices. We take into account properties such as
the lengths of attached 2-chains, i.e.~their number of vertices, and
the number of lines thereof.
In addition to these direct enhancements we iterate the ordering.
That means, having fixed order relations in a particular way,
we can impose even more ordering by taking into account order
criteria of neighbours. To this end, we define the following numbers.
For every $v\in\cB_\Gamma\setminus\cB_{\Gamma,int}^{(2)}$,
\bea
  d_1(v) & = & \sum_{w\in\cN(v)} E(w) , \\
  d_2(v) & = & \sum_{w\in\cN(v)\atop
    w\not\in\cB_{\Gamma,int}^{(2)}} N(w) , \\ \label{alg.19}
  d_3(v) & = & \sum_{w\in\cN(v)\atop
    w\not\in\cB_{\Gamma,int}^{(2)}} 1 , \\
  d_4(v) & = & \sum_{w\in\cN(v)\atop
    w\in\cB_{\Gamma,int}^{(2)}} \mu(w) , \\
  d_5(v) & = & \max_{w\in\cN(v)\atop
    w\in\cB_{\Gamma,int}^{(2)}} \mu(w) .
\eea
Then the final definition of $\cO_{ni}$ is as follows.
\be \label{alg.20}
  \cO_{ni} = < \cO_{ni,1},\ldots,\cO_{ni,5} >,
\ee
where for all $k=1,\ldots,5$ and
for all $v,w\in\cB_\Gamma\setminus\cB_{\Gamma,int}^{(2)}$,
\be \label{alg.21}
  v \geo{\cO_{ni,k}}w \; \Longleftrightarrow \;
  d_k(v) > d_k(w).
\ee

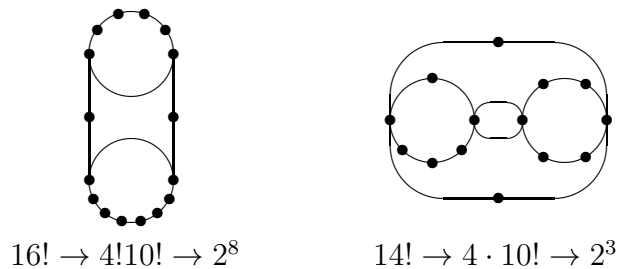
\begin{figure}[htb]
\caption{\label{perm} Two examples for the efficiency of extended
and iterated preordering. The bottom lines below the graphs show
the number of permutations left after no preordering at all, preordering
according to L\"uscher and Weisz, and extended iterated preordering.}

\begin{center}
\setlength{\unitlength}{0.8cm}
\begin{picture}(10.0,5.0)


\put(1.0,0.0){
\setlength{\unitlength}{0.8cm}
\begin{picture}(4.0,5.0)
\put(1.8,1.8){\circle{1.4}}
\put(1.1,1.8){\circle*{0.17}}
\put(2.5,1.8){\circle*{0.17}}
\put(2.4307,1.4963){\circle*{0.17}}
\put(2.2364,1.2527){\circle*{0.17}}
\put(1.9558,1.1176){\circle*{0.17}}
\put(1.6442,1.1176){\circle*{0.17}}
\put(1.3636,1.2527){\circle*{0.17}}
\put(1.1693,1.4963){\circle*{0.17}}
\put(1.1,1.8){\line(0,1){2.1}}
\put(1.1,2.85){\circle*{0.17}}
\put(2.5,1.8){\line(0,1){2.1}}
\put(2.5,2.85){\circle*{0.17}}
\put(1.8,3.9){\circle{1.4}}
\put(1.1,3.9){\circle*{0.17}}
\put(2.5,3.9){\circle*{0.17}}
\put(1.2337,4.3114){\circle*{0.17}}
\put(1.5837,4.5657){\circle*{0.17}}
\put(2.0163,4.5657){\circle*{0.17}}
\put(2.3663,4.3114){\circle*{0.17}}

\put(1.1,0.6){\makebox(1.2,0){$16!\to 4!10!\to 2^8$}}

\end{picture}
}


\put(6.0,1.00){
\setlength{\unitlength}{0.8cm}
\begin{picture}(7.0,5.0)
\put(1.8,1.8){\circle{1.4}}
\put(1.1,1.8){\circle*{0.16}}
\put(2.5,1.8){\circle*{0.16}}
\put(1.8000,2.5000){\circle*{0.16}}
\put(2.2950,1.3050){\circle*{0.16}}
\put(1.8000,1.1000){\circle*{0.16}}
\put(1.3050,1.3050){\circle*{0.16}}
\put(2.9000,1.8000){\oval(0.800,0.600)}
\put(3.3000,1.8000){\circle*{0.16}}
\put(4.0,1.8){\circle{1.4}}
\put(4.7000,1.8000){\circle*{0.16}}
\put(3.6500,2.4062){\circle*{0.16}}
\put(4.3500,2.4062){\circle*{0.16}}
\put(3.6500,1.1938){\circle*{0.16}}
\put(4.3500,1.1938){\circle*{0.16}}
\put(2.9000,1.8000){\oval(3.600,2.600)}
\put(2.9000,3.1000){\circle*{0.16}}
\put(2.9000,0.5000){\circle*{0.16}}

\put(1.1,-0.40){\makebox(3.50,0.00){$14!\to 4\cdot 10!\to 2^3$}}

\end{picture}
}

\end{picture}
\end{center}

\end{figure}

After all these technical details let us now give some examples.
To compute the canonical incidence matrix $I_\Gamma^\cO$
of a graph $\Gamma$ under consideration, the vertices
are ordered first according to the order
relation \eqn{alg.13}. In turn $I_\Gamma^\cO$ is determined by
\eqn{alg.8}. For the graph of Fig. \ref{preord} above there are
$(3!)^5$ permutations left, orders of magnitude less than 15! or
even 17!.
Two further examples are shown in Fig. \ref{perm}. Those diagrams
belong to the intermediate graph classes out of which the
"physical" graphs such as $\cS_k$ and $\cQ_k$ are constructed,
as will be discussed in the next section.


\subsection{\label{inc.ring}Ring graphs.}

A special treatment is necessary for those graphs that have internal
2-vertices only. Let $\Gamma$ be such a ring graph, so that
$\cB_\Gamma = \cB_{\Gamma,int}^{(2)}$.
The distinction between internal 2-vertices and those that are not
for non-ring graphs is replaced now by identifying so-called symmetric
2-vertices. Such a vertex is defined by the property that with both
of its neighbours it has the same number of lines in common.
More precisely, a vertex $v\in\cB_\Gamma$ is called a symmetric 2-vertex if
there is an integer $k\in{\bf N}$ such that
\be
  \nu_k(v) = 2.
\ee
The number $k$ is denoted by $c(v)$ (There is at most one such $k$ for
every vertex. For $l\not=k$, $\nu_l(v)=0$).
The set of symmetric 2-vertices is denoted by $\cB_{\Gamma,symm}$.
A ring graph $\Gamma$ is called exact if every vertex $v\in\cB_\Gamma$
is a symmetric 2-vertex.

The definitions and constructions of 2-chains, order relations on them and
subsequently on internal 2-vertices as discussed in the last section
for non-ring graphs can now be translated to ring graphs. The essential step
is to replace the notion of internal 2-vertices by symmetric
2-vertices, and similarly for their complements. Some of the notions
used to define the order relations above have to be adjusted
correspondingly, other ones become dummy.
The details are very similar as in the last section. For completeness
they are given in the Appendix for non-exact ring graphs.

We are left to define the canonical incidence matrix for exact ring graphs.
This is done explicitly. Let $\Gamma$ be an exact ring graph with
$L$ lines and $V$ vertices. Then $V\geq 3$, and $\nu=L/V$ is a positive
integer. The canonical incidence matrix $\widehat I_\Gamma$ is defined as
the symmetric $V\times V$ matrix with the only non-vanishing entries of its
right upper triangle being given as follows.
\bea
  \widehat I_\Gamma (k,k+2) &=&\nu  \; ,\;\mbox{for}
   \; k=1,\ldots,V-2 ,\\
  \widehat I_\Gamma (1,2) = \widehat I_\Gamma (V-1,V) &=&\nu .
\eea
In particular, the diagonal entries are zero. The symmetry
number is given by $S(\Gamma)=2V \cdot (\nu !)^V$.


\section{Generation of graphs and weight computation.}

In this section we present and discuss efficient algorithms to generate
all the equivalence classes of graphs we are looking for
to compute correlation functions and susceptibilities.
The central idea is to generate them from base classes of diagrams
that are simpler in structure and less in number.
These classes should be universal in the sense that the various
correlations are generated from the same base classes. The latter
have to be provided just once.
The actual order by order construction of graphs on the number of
lines is essentially done in the simplest class only.

\subsection{Inheritance tree of graph classes}

In Sect.\ref{basic.0} we have outlined the way connected correlations
and susceptibilities can be built up from 1PI ones.
The latter allow for some kind of product representation by
1VI graphs $\cS_k$ and renormalized moments $\cQ_k$,
(\ref{lce.31}-\ref{lce.32}). Only those graphs have to be constructed
explicitly. The most apparent
difference between graphs of $\cS_k$ and $\cQ_k$ is the way they
have external lines attached.
Every $\Gamma\in\cS_k$ has at least two external vertices, whereas all
external lines of any $\Gamma\in\cQ_k$ enter the same vertex.

We generate those diagrams from general 1PI ones with internal lines only
by attaching external lines in the various different ways. This is
possible because the 1PI property is independent of external lines.
Even further simplification arise if we omit multiple lines between
vertices. Any 1PI graph with multiple lines can be constructed from
a connected graph with single lines only, by multiplying internal lines
appropriately. Actually, care is needed here because we have to relax
the 1PI property for those simplest graphs.

Fortunately, we can work around this problem. We allow for double lines
under special circumstances, still having an acceptibly small base class.
This design frees us from having to check 1PI properties for more
advanced classes during the construction process.
Also, graphs do not need to be glued together.
We are now going to state our design of graph classes and the sequence of
construction mechanism.

Although the description here is in terms of graphs themselves, the
implementation on the computer is done in terms of incidence matrices
and is more or less straight forward.
Let
\[
  \Gamma = (\cL_\Gamma,\cB_\Gamma,E_\Gamma,\Phi_\Gamma),
\]
be a diagram and $\cM\subseteq\cL_\Gamma$ a subset of the lines of
$\Gamma$. Then we set
\be
  \Gamma\setminus\cM =
   (\overline\cL_\Gamma,\cB_\Gamma,E_\Gamma,\overline\Phi_\Gamma),
\ee
where $\overline\cL_\Gamma = \cL_\Gamma\setminus\cM$ and
$\overline\Phi_\Gamma$ is the incidence relation $\Phi_\Gamma$
restricted to $\overline\Phi_\Gamma$,
\[
  \overline\Phi_\Gamma =
   \left. \Phi_\Gamma \right\vert_{\overline\cL_\Gamma} .
\]
$\Gamma\setminus\cM$ is the graph $\Gamma$ with the lines of $\cM$
omitted.
With this notion, the base classes $\cP_1(L)$ for the various numbers
of lines $L$ are defined as follows.

$\cP_1(L)$ is the set of graphs $\Gamma$ that have $L$ internal lines,
no external lines, and are 1PI, i.e.
\be
  \Gamma\in\cG_0^{\rm 1PI}(L).
\ee
Furthermore, for all pairs of vertices $v,w\in\cB_\Gamma$,
either there is at most one line between $v$ and $w$, $m(v,w)\leq 1$,
or $m(v,w)=2$ and $\Gamma\setminus\cM(v,w)$ is not connected.
Here, $\cM(v,w)$ is the set of lines between $v$ and $w$,
as defined by \eqn{lce.20}.


\begin{figure}[htb]
\caption{\label{class} The structure of the graph class $\cP_1$.
The hatched bubbles represent the reduced components, which are
connected by double lines.}

\begin{center}
\setlength{\unitlength}{0.5cm}
\begin{picture}(8.0,5.5)


\put(4.0,4.0){\circle{1.4}}
\put(3.3040,4.075){\line(1,0){1.3919}}
\put(3.3040,3.925){\line(1,0){1.3919}}
\put(3.3371,4.225){\line(1,0){1.3257}}
\put(3.3371,3.775){\line(1,0){1.3257}}
\put(3.4089,4.375){\line(1,0){1.1822}}
\put(3.4089,3.625){\line(1,0){1.1822}}
\put(3.5370,4.525){\line(1,0){0.9260}}
\put(3.5370,3.475){\line(1,0){0.9260}}
\put(3.8146,4.675){\line(1,0){0.3708}}
\put(3.8146,3.325){\line(1,0){0.3708}}
\put(4.075,3.3040){\line(0,1){1.3919}}
\put(3.925,3.3040){\line(0,1){1.3919}}
\put(4.225,3.3371){\line(0,1){1.3257}}
\put(3.775,3.3371){\line(0,1){1.3257}}
\put(4.375,3.4089){\line(0,1){1.1822}}
\put(3.625,3.4089){\line(0,1){1.1822}}
\put(4.525,3.5370){\line(0,1){0.9260}}
\put(3.475,3.5370){\line(0,1){0.9260}}
\put(4.675,3.8146){\line(0,1){0.3708}}
\put(3.325,3.8146){\line(0,1){0.3708}}
\put(2.900,4.000){\oval(0.800,0.600)}
\put(2.500,4.000){\circle*{0.20}}
\put(3.300,4.000){\circle*{0.20}}
\put(1.8,4.0){\circle{1.4}}
\put(1.1040,4.075){\line(1,0){1.3919}}
\put(1.1040,3.925){\line(1,0){1.3919}}
\put(1.1371,4.225){\line(1,0){1.3257}}
\put(1.1371,3.775){\line(1,0){1.3257}}
\put(1.2089,4.375){\line(1,0){1.1822}}
\put(1.2089,3.625){\line(1,0){1.1822}}
\put(1.3370,4.525){\line(1,0){0.9260}}
\put(1.3370,3.475){\line(1,0){0.9260}}
\put(1.6146,4.675){\line(1,0){0.3708}}
\put(1.6146,3.325){\line(1,0){0.3708}}
\put(1.875,3.3040){\line(0,1){1.3919}}
\put(1.725,3.3040){\line(0,1){1.3919}}
\put(2.025,3.3371){\line(0,1){1.3257}}
\put(1.575,3.3371){\line(0,1){1.3257}}
\put(2.175,3.4089){\line(0,1){1.1822}}
\put(1.425,3.4089){\line(0,1){1.1822}}
\put(2.325,3.5370){\line(0,1){0.9260}}
\put(1.275,3.5370){\line(0,1){0.9260}}
\put(2.475,3.8146){\line(0,1){0.3708}}
\put(1.125,3.8146){\line(0,1){0.3708}}
\put(5.100,4.000){\oval(0.800,0.600)}
\put(4.700,4.000){\circle*{0.20}}
\put(5.500,4.000){\circle*{0.20}}
\put(6.2,4.0){\circle{1.4}}
\put(5.5040,4.075){\line(1,0){1.3919}}
\put(5.5040,3.925){\line(1,0){1.3919}}
\put(5.5371,4.225){\line(1,0){1.3257}}
\put(5.5371,3.775){\line(1,0){1.3257}}
\put(5.6089,4.375){\line(1,0){1.1822}}
\put(5.6089,3.625){\line(1,0){1.1822}}
\put(5.7370,4.525){\line(1,0){0.9260}}
\put(5.7370,3.475){\line(1,0){0.9260}}
\put(6.0146,4.675){\line(1,0){0.3708}}
\put(6.0146,3.325){\line(1,0){0.3708}}
\put(6.275,3.3040){\line(0,1){1.3919}}
\put(6.125,3.3040){\line(0,1){1.3919}}
\put(6.425,3.3371){\line(0,1){1.3257}}
\put(5.975,3.3371){\line(0,1){1.3257}}
\put(6.575,3.4089){\line(0,1){1.1822}}
\put(5.825,3.4089){\line(0,1){1.1822}}
\put(6.725,3.5370){\line(0,1){0.9260}}
\put(5.675,3.5370){\line(0,1){0.9260}}
\put(6.875,3.8146){\line(0,1){0.3708}}
\put(5.525,3.8146){\line(0,1){0.3708}}
\put(7.300,4.000){\oval(0.800,0.600)}
\put(6.900,4.000){\circle*{0.20}}
\put(7.700,4.000){\circle*{0.20}}
\put(8.4,4.0){\circle{1.4}}
\put(7.7040,4.075){\line(1,0){1.3919}}
\put(7.7040,3.925){\line(1,0){1.3919}}
\put(7.7371,4.225){\line(1,0){1.3257}}
\put(7.7371,3.775){\line(1,0){1.3257}}
\put(7.8089,4.375){\line(1,0){1.1822}}
\put(7.8089,3.625){\line(1,0){1.1822}}
\put(7.9370,4.525){\line(1,0){0.9260}}
\put(7.9370,3.475){\line(1,0){0.9260}}
\put(8.2146,4.675){\line(1,0){0.3708}}
\put(8.2146,3.325){\line(1,0){0.3708}}
\put(8.475,3.3040){\line(0,1){1.3919}}
\put(8.325,3.3040){\line(0,1){1.3919}}
\put(8.625,3.3371){\line(0,1){1.3257}}
\put(8.175,3.3371){\line(0,1){1.3257}}
\put(8.775,3.4089){\line(0,1){1.1822}}
\put(8.025,3.4089){\line(0,1){1.1822}}
\put(8.925,3.5370){\line(0,1){0.9260}}
\put(7.875,3.5370){\line(0,1){0.9260}}
\put(9.075,3.8146){\line(0,1){0.3708}}
\put(7.725,3.8146){\line(0,1){0.3708}}
\put(6.200,2.900){\oval(0.600,0.800)}
\put(6.200,3.300){\circle*{0.20}}
\put(6.200,2.500){\circle*{0.20}}
\put(6.2,1.8){\circle{1.4}}
\put(5.5040,1.875){\line(1,0){1.3919}}
\put(5.5040,1.725){\line(1,0){1.3919}}
\put(5.5371,2.025){\line(1,0){1.3257}}
\put(5.5371,1.575){\line(1,0){1.3257}}
\put(5.6089,2.175){\line(1,0){1.1822}}
\put(5.6089,1.425){\line(1,0){1.1822}}
\put(5.7370,2.325){\line(1,0){0.9260}}
\put(5.7370,1.275){\line(1,0){0.9260}}
\put(6.0146,2.475){\line(1,0){0.3708}}
\put(6.0146,1.125){\line(1,0){0.3708}}
\put(6.275,1.1040){\line(0,1){1.3919}}
\put(6.125,1.1040){\line(0,1){1.3919}}
\put(6.425,1.1371){\line(0,1){1.3257}}
\put(5.975,1.1371){\line(0,1){1.3257}}
\put(6.575,1.2089){\line(0,1){1.1822}}
\put(5.825,1.2089){\line(0,1){1.1822}}
\put(6.725,1.3370){\line(0,1){0.9260}}
\put(5.675,1.3370){\line(0,1){0.9260}}
\put(6.875,1.6146){\line(0,1){0.3708}}
\put(5.525,1.6146){\line(0,1){0.3708}}
\put(7.300,1.800){\oval(0.800,0.600)}
\put(6.900,1.800){\circle*{0.20}}
\put(7.700,1.800){\circle*{0.20}}
\put(8.4,1.8){\circle{1.4}}
\put(7.7040,1.875){\line(1,0){1.3919}}
\put(7.7040,1.725){\line(1,0){1.3919}}
\put(7.7371,2.025){\line(1,0){1.3257}}
\put(7.7371,1.575){\line(1,0){1.3257}}
\put(7.8089,2.175){\line(1,0){1.1822}}
\put(7.8089,1.425){\line(1,0){1.1822}}
\put(7.9370,2.325){\line(1,0){0.9260}}
\put(7.9370,1.275){\line(1,0){0.9260}}
\put(8.2146,2.475){\line(1,0){0.3708}}
\put(8.2146,1.125){\line(1,0){0.3708}}
\put(8.475,1.1040){\line(0,1){1.3919}}
\put(8.325,1.1040){\line(0,1){1.3919}}
\put(8.625,1.1371){\line(0,1){1.3257}}
\put(8.175,1.1371){\line(0,1){1.3257}}
\put(8.775,1.2089){\line(0,1){1.1822}}
\put(8.025,1.2089){\line(0,1){1.1822}}
\put(8.925,1.3370){\line(0,1){0.9260}}
\put(7.875,1.3370){\line(0,1){0.9260}}
\put(9.075,1.6146){\line(0,1){0.3708}}
\put(7.725,1.6146){\line(0,1){0.3708}}
\put(5.100,1.800){\oval(0.800,0.600)}
\put(4.700,1.800){\circle*{0.20}}
\put(5.500,1.800){\circle*{0.20}}
\put(4.0,1.8){\circle{1.4}}
\put(3.3040,1.875){\line(1,0){1.3919}}
\put(3.3040,1.725){\line(1,0){1.3919}}
\put(3.3371,2.025){\line(1,0){1.3257}}
\put(3.3371,1.575){\line(1,0){1.3257}}
\put(3.4089,2.175){\line(1,0){1.1822}}
\put(3.4089,1.425){\line(1,0){1.1822}}
\put(3.5370,2.325){\line(1,0){0.9260}}
\put(3.5370,1.275){\line(1,0){0.9260}}
\put(3.8146,2.475){\line(1,0){0.3708}}
\put(3.8146,1.125){\line(1,0){0.3708}}
\put(4.075,1.1040){\line(0,1){1.3919}}
\put(3.925,1.1040){\line(0,1){1.3919}}
\put(4.225,1.1371){\line(0,1){1.3257}}
\put(3.775,1.1371){\line(0,1){1.3257}}
\put(4.375,1.2089){\line(0,1){1.1822}}
\put(3.625,1.2089){\line(0,1){1.1822}}
\put(4.525,1.3370){\line(0,1){0.9260}}
\put(3.475,1.3370){\line(0,1){0.9260}}
\put(4.675,1.6146){\line(0,1){0.3708}}
\put(3.325,1.6146){\line(0,1){0.3708}}
\end{picture}
\end{center}

\end{figure}
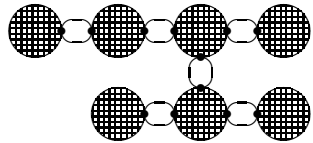

Thus, removing any one of a double line between two vertices destroys
the 1PI property, removing both disconnects the graph.
Fig. \ref{class} shows the typical structure of a graph
$\Gamma\in\cP_1(L)$. Removing all double lines, $\Gamma$ decomposes
into various connected components. The latter are 1PI, and any
two vertices have at most one line in common. Those parts are called
reduced components. Replacing each reduced component by a single
vertex, and every double line of $\Gamma$ by a single line, the
resulting graph is a tree, i.e.~it does not have any loop.

{}From the $\cP_1$ we derive the next more involved and larger classes
$\cP_2(L)$ which still lacks external lines, but have an arbitrary
number of lines between their vertices,
\be
  \cP_2(L) = \cG_0^{\rm 1PI}.
\ee
All the graphs of $\cP_2(L)$ are obtained by taking every graph of
$\cP_1(L^\prime)$ and multiplying its lines in all possible ways,
so that in total $L-L^\prime$ lines are added. This has to be done
for every $2\leq L^\prime\leq L$ whenever $L>0$.
In this way, for every equivalence class of $\cP_2(L)$ there will be
at least one graph constructed, i.e.~$\cP_2(L)$ is covered.
However, many graphs generated in this way are equivalent, and we have
to keep track that for every equivalence class only one
representant is kept.
Fortunately, we don't have to compare a newly generated graph to all prior
constructed diagrams. This is due to the fact that two graphs of
$\cP_2(L)$ can be equivalent only if they are both generated from the
same graph of some $\cP_1(L^\prime)$ according to the above rules.

The last step of the graph generation sequence is to generate the
1VI graphs $\cS_k(L)$ and the renormalized moments $\cQ_k(L)$ out of
the graphs of $\cP_2(L)$. This is done by fetching all graphs of
$\cP_2(L)$, and attaching $k$ external lines in different ways.
Attaching all of them to just one vertex generates $Q_k(L)$.

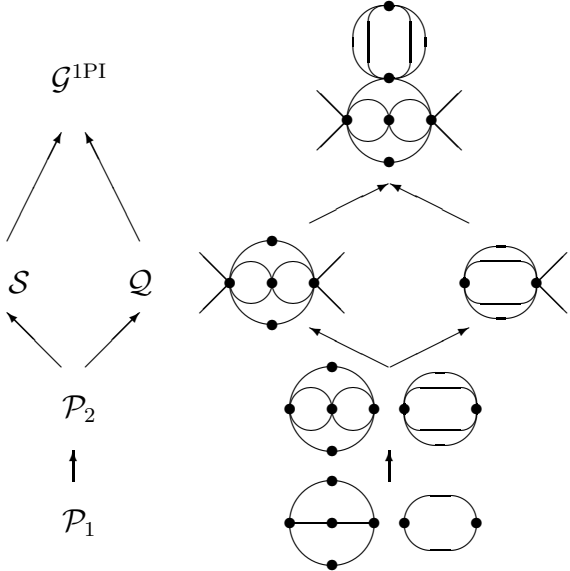
\begin{figure}[htb]
\caption{\label{inherit} Inheritance tree of graph classes
$\cP_1\to\cP_2\to\cS$ and $\cQ$.
The right hand side of the figure shows an example construction.
Actually, the graphs of $\cG^{\rm 1PI}$ are no more constructed
explicitly.}

\begin{center}
\setlength{\unitlength}{0.8cm}
\begin{picture}(10.0,10.0)


\put(1.0,0.0){
\setlength{\unitlength}{0.8cm}
\begin{picture}(2.0,10.0)
\put(1.0,0.8){\makebox(1.2,0){$\cP_1$}}
\put(1.5,1.5){\vector(0,1){0.5}}
\put(1.0,2.7){\makebox(1.2,0){$\cP_2$}}
\put(1.3,3.4){\vector(-1,1){0.9}}
\put(1.7,3.4){\vector(1,1){0.9}}
\put(0.0,4.8){\makebox(1.2,0){$\cS$}}
\put(2.0,4.8){\makebox(1.2,0){$\cQ$}}
\put(0.4,5.5){\vector(1,2){0.9}}
\put(2.6,5.5){\vector(-1,2){0.9}}
\put(1.0,8.2){\makebox(1.2,0){$\cG^{\rm 1PI}$}}

\end{picture}
}


\put(5.0,0.0){
\setlength{\unitlength}{0.8cm}
\begin{picture}(7.0,10.0)
\put(1.8,0.8){\circle{1.4}}
\put(1.1,0.8){\line(1,0){1.4}}
\put(1.1,0.8){\circle*{0.16}}
\put(2.5,0.8){\circle*{0.16}}
\put(1.8,1.5){\circle*{0.16}}
\put(1.8,0.1){\circle*{0.16}}
\put(1.8,0.8){\circle*{0.16}}
\put(3.6,0.8){\oval(1.2,0.9)}
\put(3.0,0.8){\circle*{0.16}}
\put(4.2,0.8){\circle*{0.16}}
\put(2.75,1.5){\vector(0,1){0.5}}
\put(1.8,2.7){\circle{1.4}}
\put(1.45,2.7){\circle{0.7}}
\put(2.15,2.7){\circle{0.7}}
\put(1.8,2.7){\circle*{0.16}}
\put(1.1,2.7){\circle*{0.16}}
\put(2.5,2.7){\circle*{0.16}}
\put(1.8,3.4){\circle*{0.16}}
\put(1.8,2.0){\circle*{0.16}}
\put(3.6,2.7){\oval(1.2,0.7)}
\put(3.6,2.7){\oval(1.2,1.2)}
\put(3.0,2.7){\circle*{0.16}}
\put(4.2,2.7){\circle*{0.16}}
\put(2.73,3.4){\vector(-2,1){1.3}}
\put(2.77,3.4){\vector(2,1){1.3}}
\put(0.8,4.8){\circle{1.4}}
\put(0.45,4.8){\circle{0.7}}
\put(1.15,4.8){\circle{0.7}}
\put(0.8,4.8){\circle*{0.16}}
\put(0.1,4.8){\circle*{0.16}}
\put(1.5,4.8){\circle*{0.16}}
\put(0.8,5.5){\circle*{0.16}}
\put(0.8,4.1){\circle*{0.16}}
\put(1.5,4.8){\line(1,1){0.5}}
\put(1.5,4.8){\line(1,-1){0.5}}
\put(0.1,4.8){\line(-1,1){0.5}}
\put(0.1,4.8){\line(-1,-1){0.5}}
\put(4.6,4.8){\oval(1.2,0.7)}
\put(4.6,4.8){\oval(1.2,1.2)}
\put(4.0,4.8){\circle*{0.16}}
\put(5.2,4.8){\circle*{0.16}}
\put(5.2,4.8){\line(1,1){0.5}}
\put(5.2,4.8){\line(1,-1){0.5}}
\put(1.43,5.8){\vector(2,1){1.3}}
\put(4.07,5.8){\vector(-2,1){1.3}}
\put(2.75,7.5){\circle{1.4}}
\put(2.40,7.5){\circle{0.7}}
\put(3.10,7.5){\circle{0.7}}
\put(2.75,7.5){\circle*{0.16}}
\put(2.05,7.5){\circle*{0.16}}
\put(3.45,7.5){\circle*{0.16}}
\put(2.75,8.2){\circle*{0.16}}
\put(2.75,6.8){\circle*{0.16}}
\put(3.45,7.5){\line(1,1){0.5}}
\put(3.45,7.5){\line(1,-1){0.5}}
\put(2.05,7.5){\line(-1,1){0.5}}
\put(2.05,7.5){\line(-1,-1){0.5}}
\put(2.75,8.8){\oval(0.7,1.2)}
\put(2.75,8.8){\oval(1.2,1.2)}
\put(2.75,9.4){\circle*{0.16}}

\end{picture}
}

\end{picture}
\end{center}

\end{figure}

The graphs of $\cS_k(L)$ have at least two external vertices once $L>0$.
Furthermore, they have to be 1VI, which restricts the number of
possibilities to attach the external lines.
To check this property needs one recursive depths first search
\cite{sedgewick} within the augmented graph. It is obtained by adding
one vertex and attaching all external lines to it \cite{TR1,lowenstein}.
This is a very cheap algorithm and can be done before the incidence
matrix of the graph is brought to its canonical form.
A further simplification arises for the O(N) scalar models due to the
fact that we can restrict $\cS_k$ and $\cQ_k$ to even graphs only.

Again, only graphs that are mutually non-equivalent have to be kept.
Similar as before, two graphs can be equivalent only if they originate from
the same graph of $\cP_2(L)$.

The generation sequence of the classes is scetched in Fig. \ref{inherit}.

\begin{figure}[htb]
\caption{\label{p1class} The graphs of $P_1(L)$ for all $L\leq 4$.}

\begin{center}
\setlength{\unitlength}{0.8cm}
\begin{picture}(10.0,3.0)


\put(0.0,0.0){
\setlength{\unitlength}{0.8cm}
\begin{picture}(2.5,5.0)

\put(1.25,1.0){\circle*{0.16}}

\end{picture}
}


\put(2.0,0.0){
\setlength{\unitlength}{0.8cm}
\begin{picture}(2.5,5.0)

\put(1.25,1.0){\circle{1.0}}
\put(0.75,1.0){\circle*{0.16}}
\put(1.75,1.0){\circle*{0.16}}

\end{picture}
}


\put(4.75,0.0){
\setlength{\unitlength}{0.8cm}
\begin{picture}(2.5,5.0)

\put(0.75,1.0){\circle{1.0}}
\put(1.75,1.0){\circle{1.0}}
\put(0.25,1.0){\circle*{0.16}}
\put(1.25,1.0){\circle*{0.16}}
\put(2.25,1.0){\circle*{0.16}}

\end{picture}
}


\put(7.5,0.0){
\setlength{\unitlength}{0.8cm}
\begin{picture}(2.5,5.0)

\put(1.25,1.0){\circle{1.4}}
\put(0.55,1.0){\circle*{0.16}}
\put(1.95,1.0){\circle*{0.16}}
\put(1.25,1.7){\circle*{0.16}}
\put(1.25,0.3){\circle*{0.16}}

\end{picture}
}

\end{picture}
\end{center}

\end{figure}
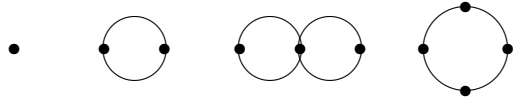

It remains to construct the base classes $\cP_1(L)$.
This is done inductively on $L$. We closely follow the proposal
of \cite{LW1}, adjusted to the present case.
The lowest order graphs are given explicitly, $\cP_1(L)$ for $L\leq 4$,
as in Fig. \ref{p1class}.

Now suppose that $L\geq 5$. Then $\cP_1(L)$ is constructed from
$\cP_1(L^\prime)$ with $L^\prime = L-1,L-2$ by three rules,
\be
  \cP_1(L-1)\;\stackrel{R_1}{\to}\;\cP_1(L) \quad , \quad
  \cP_1(L-2)\;\stackrel{R_2,R_3}{\to}\;\cP_1(L) .
\ee
They apply to every graph of the respective class and operate separately
on every pair of vertices that satisfy appropriate conditions.
To state these rules we have to provide some definitions first.

Let $\Gamma\in\cP_1(L)$ for some $L$. Then to every vertex
$v\in\cB_\Gamma$ we assign

\begin{itemize}
\item a parity number $\eta(v)\in\{-1,+1\}$ in such a way that neighboured
vertices have opposite parity. This is unique up to a global sign for
$\Gamma$, due to the fact that every loop of $\Gamma$ has an even
number of lines.
\item a reduced component index $r(v)\in\{1,2,\ldots,n\}$, where
$n$ denotes the number of reduced components of $\Gamma$. This is done
in such a way that vertices that belong to the same reduced component
get the same number $r$ attached. Those numbers are unique up to
permutations of $\{1,\ldots,n\}$.
\end{itemize}

With these notions, the rules operate as follows.

\begin{description}
\item[$R_1$] on $\Gamma\in\cP_1(L-1)$ and for $v_1,v_2\in\cB_\Gamma$ with
\hfill\newline
$\eta(v_1)\not=\eta(v_2)$, $m(v_1,v_2)=0$,
$r(v_1)=r(v_2)$,
\hfill\newline
insert a line between $v_1$ and $v_2$,
\be
  \begin{array}{c} {\bullet\mbox{$\qquad\qquad$}\bullet} \\
         {v_1\mbox{$\qquad\qquad$} v_2} \end{array}
  \quad\longrightarrow\quad
  \begin{array}{c} {\bullet\!\mbox{------------}\!\bullet} \\
         {v_1\mbox{$\qquad\qquad$} v_2} \end{array} .
\ee
In particular, $v_1\not= v_2$.
\item[$R_2$] on $\Gamma\in\cP_1(L-2)$ and for $v_1,v_2\in\cB_\Gamma$ with
\hfill\newline
$\eta(v_1)=\eta(v_2)$ (so that $m(v_1,v_2)=0$),
$r(v_1)=r(v_2)$,
\hfill\newline
insert a vertex $w$ and 2 single lines connecting $w$ to $v_1$ and to $v_2$,
\be
  \begin{array}{c} {\bullet\mbox{$\qquad\qquad$}\bullet} \\
        {v_1\mbox{$\qquad\qquad$} v_2} \end{array}
  \quad\longrightarrow\quad
  \begin{array}{c}
        {\bullet\!\mbox{--------}\!\!\bullet\!\mbox{--------}\!\bullet}
      \\  {v_1\mbox{$\qquad$} w \mbox{$\qquad$} v_2} \end{array}.
\ee
There is no line and there will be no line between $v_1$ and $v_2$.
\item[$R_3$] on $\Gamma\in\cP_1(L-2)$ and for $v_1,v_2\in\cB_\Gamma$ with
\hfill\newline
$\cM(v_1,v_2)\not=\emptyset$, and at least one vertex has more than 2
lines attached.
\hfill\newline
Let $l\in\cM(v_1,v_2)$. This line $l$ is replaced by
2 vertices $w_1,w_2$ and 3 new lines forming a 2-chain
$(v_1,w_1,w_2,v_2)$ of single lines,
\be
  \begin{array}{c} {\bullet\!\mbox{-----------}\!\bullet} \\
        {v_1\mbox{$\qquad\qquad$} v_2} \end{array}
  \quad\longrightarrow\quad
  \begin{array}{c}
        {\bullet\!\mbox{--------}\!\!\bullet\!\mbox{--------}\!\!\bullet
     \!\mbox{--------}\!\bullet} \\
       {v_1\mbox{$\qquad$} w_1 \mbox{$\qquad$} w_2
        \mbox{$\qquad$} v_2}  \end{array}.
\ee
\end{description}

Some remarks are in order. For the first two rules, both vertices must
belong to the same reduced component, otherwise the class $\cP_1$
would be left. In the 2.~case, the two vertices $v_1$ and $v_2$ can
be equal. In this case a double line is generated, and a new reduced
component appears, consisting of a single vertex only.

\begin{figure}[htb]
\caption{\label{oponchain} To a long 2-chain, rule R3 need to be
applied only once.}

\begin{center}
\setlength{\unitlength}{0.8cm}
\begin{picture}(10.0,3.0)


\put(0.0,0.0){
\setlength{\unitlength}{0.8cm}
\begin{picture}(3.0,3.0)

\put(0.0,2.0){\circle*{0.16}}
\put(0.0,2.0){\line(1,0){2.0}}
\put(0.0,2.0){\line(-1,1){0.5}}
\put(0.0,2.0){\line(-1,-1){0.5}}
\put(0.5,2.0){\circle*{0.16}}
\put(1.0,2.0){\circle*{0.16}}
\put(1.5,2.0){\circle*{0.16}}
\put(2.0,2.0){\circle*{0.16}}
\put(2.0,2.0){\line(1,1){0.5}}
\put(2.0,2.0){\line(1,-1){0.5}}

\end{picture}
}


\put(2.8,0.0){
\setlength{\unitlength}{0.8cm}
\begin{picture}(0.4,3.0)

\put(0.0,2.0){\makebox(0.4,0){$\to$}}

\end{picture}
}


\put(4.0,0.0){
\setlength{\unitlength}{0.8cm}
\begin{picture}(4.0,3.0)

\put(0.0,2.0){\circle*{0.16}}
\put(0.0,2.0){\line(1,0){3.0}}
\put(0.0,2.0){\line(-1,1){0.5}}
\put(0.0,2.0){\line(-1,-1){0.5}}
\put(0.5,2.0){\circle*{0.16}}
\put(1.0,2.0){\circle*{0.16}}
\put(1.5,2.0){\circle*{0.16}}
\put(2.0,2.0){\circle*{0.16}}
\put(2.5,2.0){\circle*{0.16}}
\put(3.0,2.0){\circle*{0.16}}
\put(3.0,2.0){\line(1,1){0.5}}
\put(3.0,2.0){\line(1,-1){0.5}}

\end{picture}
}

\end{picture}
\end{center}

\end{figure}
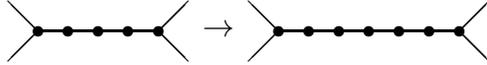


The 3.~rule applies also if the line $l$ belongs to a double line.
Obviously, we need to operate only on one of those double lines. Two
reduced components melt to a single one in this case.
Furthermore, for a 2-chain as in Fig. \ref{oponchain}, it is sufficient
to operate on the left or right end.

Again we have to check that every newly generated graph provides a
new equivalence class. In contrast to the construction of
$\cP_2$, $\cS$ and $\cQ$ discussed above, we have to compare to all the
graphs generated yet (with same number of lines and vertices, of course).
But this number is considerably smaller than that of the more involved
classes.

Definition and construction mechanism of the intermediate classes
is not unique, of course. Our choice and design of the classes differ
from those of L\"uscher and Weisz \cite{LW1}.
We allow graphs of the base class $\cP_1$ not to be twice connected.
This increases its number of graphs, but it is still acceptibly small.
The major advantage of our approach is that we do not have to glue
graphs in any way.
The costs increase with minor power with respect to the number of graphs.

\subsection{Weight computation.}

Every graph represents a real number it contributes to a susceptibility,
its weight. For the O(N) scalar models the most laborious
parts of it are the computation of both the lattice imbedding number and
the O(N) factors.
The other contributing factors are more easily obtained.
In particular, the topological symmetry number is computed in the
course of the graph generation.

The lattice imbedding number of a connected graph $\Gamma$
is given by the number of possibilities all its vertices $v\in\cB_\Gamma$
can be placed at lattice sites $x(v)$
consistent with the topology of $\Gamma$.
That is, vertices connected by a line have to be placed at nearest
neighbour lattice sites. One vertex can be fixed, saving the extensive
volume factor. It is the nature of the LCE that no exclusion principle
applies, i.e.~different vertices can be placed onto the same lattice site.

For the explicit computation of the lattice imbedding number we use
the techniques proposed in \cite{LW1} appropriately generalized.
The central idea is to place only those vertices $v$ explicitly that are
not internal 2-vertices, $v\not\in\cB_{\Gamma,int}^{(2)}$.
They are connected by maximal 2-chains. The
latter are considered as representing random walks of fixed length.
Once all $v\not\in\cB_{\Gamma,int}^{(2)}$ are put onto the lattice,
every 2-chain contributes a factor given by the number of random walks
of its length from one end point to the other.
Closed analytic expressions can be derived for them.
In turn, summation is done over all placements of the
$v\not\in\cB_{\Gamma,int}^{(2)}$.
The shortest 2-chain between two vertices determines the maximal
distance of the lattice sites they can be placed at.
On the $L_0\times\infty^{D-1}$ lattice $\Lambda_1$,
the distance is defined by
\be
  (x,y) = || x_0-y_0 ||_{L_0} + \sum_{i=1}^{D-1} |x_i-y_i|
   \quad\mbox{for}\; x,y\in\Lambda_1 ,
\ee
where
\be
  || x_0 ||_{L_0} = \inf_{n\in{\bf Z}} | x_0+n L_0 | \leq \frac{L_0}{2}.
\ee
To avoid unefficient trial and error,
placing the vertices $v\not\in\cB_{\Gamma,int}^{(2)}$
on the lattice is done in fixed order
determined by the so-called imbedding preordering.
This enumeration is defined as follows.
The only vertex chosen to be placed at some fixed lattice site
is attached the smallest number 1.
All the vertices $v\not\in\cB_{\Gamma,int}^{(2)}$ attached to it by
a 2-chain, say $m$ in number,
get the numbers $2,\ldots,m+1$ attached, in arbitrary order.
In this way we proceed iteratively. Having enumerated a subset of
vertices, enumeration is continued with those vertices
$v\not\in\cB_{\Gamma,int}^{(2)}$ that as yet have not been attached
an imbedding preorder number, and that are connected to the subset
by a 2-chain. This is done until all $v\not\in\cB_{\Gamma,int}^{(2)}$
have got a number. One breadth first search is necessary to obtain
this enumeration.
The actual placement onto the lattice is done in the order of increasing
imbedding preorder number.

Connectivities have to be respected. This is best done by the imbedding
preordered incidence matrix $I_\Gamma^{\rm B}$. Its $(i,j)$th entry
$I_\Gamma^{\rm B}(i,j)$ is given by the length of the shortest 2-chain
between the vertices $v_i$ and $v_j$, both
$v_i,v_j\not\in\cB_{\Gamma,int}^{(2)}$, with respect to the
enumeration just defined.
In contrast to the graph construction,
we do not have to worry about uniqueness.

Imbedding the graph onto the lattice now proceeds by sweeping over the
right upper triangle of $I_\Gamma^{\rm B}$ from left to right, top to
bottom. Entering a non-zero entry $I_\Gamma^{\rm B}(k,l)$ implies
either that $v_l$ has to be placed onto the lattice, or if placed
already, verify that
\be
  ( x(v_k),x(v_l) ) \leq I_\Gamma^{\rm B}(k,l).
\ee
If satisfied, we continue. Otherwise, we re-place $v_l$ within the
appropriate cube
\be
  ( x(v_l),x(v_{l^\prime}) ) \leq I_\Gamma^{\rm B}(l,l^\prime),
\ee
where $l^\prime$ is the smalllest imbedding preorder number
such that $v_l$ and $v_{l^\prime}$ are connected by a 2-chain. At the same
time, all $v_m$ with $m>l$ are put off the lattice.
A succesful sweep implies that a particular imbedding is found.
All the random walk numbers of the 2-chains are multiplied together.
Imbedding search continues by moving the latest vertex that can be placed
to a position not yet taken within its cube, and proceed as before.
If it is the first (fixed) vertex, we are done.

The discussion above concerns the computation of the lattice imbedding
number of a single connected graph.
Actually, for classes of graphs we don't need to compute all those
numbers separately. This is because imbedding numbers do not depend
on multiple lines between vertices.
Only the nearest neighbour constraint matters.
Replacing multiple lines by single lines, many
graphs become topologically equivalent and thus have identical imbedding
number. To make this working needs canonical ordering and buffering.
Further improvement is obtained if no explicit weight factors such
as $g(x,y)$, (\ref{lce.10}f), are involved. In this case we
do not need to distinguish between internal and external vertices.

The way we have done
the computation of the O(N) factors of a graph $\Gamma$, i.e.~the
summation over all O(N) symmetry labels attached to the internal lines
of $\Gamma$, proceeds as follows.
All the external lines are  connected in pairs by (possibly trivial)
paths of lines through the graph.
Once all the external lines are paired in a particular way, all
the remaining
internal lines have to be combined to loops in all possible ways.
Every succesfull covering of the lines yield a $N^M$, where $M$
is the number of loops. The sum over all those configurations yields
the O(N) factor of $\Gamma$.

Let us make the notions more precise. First of all, to avoid overcounting,
we enumerate the external lines.
Let $\Gamma = (\cL_\Gamma,\cB_\Gamma,E_\Gamma,\Phi_\Gamma)$ be a graph,
$L$ and $E=\sum_{v\in\cB_\Gamma}E_\Gamma(v)$ its number of internal and
external lines, respectively.
An enumeration of the external lines of $\Gamma$ is a map
\be
  \eta: \{ 1,\ldots,E\} \to \cB_{\Gamma,ext}
\ee
such that for every $v\in\cB_{\Gamma,ext}$, the set
\be
  \{ i\in\{1,\ldots,E\} \; \vert \; \eta(i)=v \}
\ee
has precisely $E_\Gamma(v)$ elements.
In practice, enumeration is done in conformity with canonical ordering.
That is, for $i<j$, $\eta(i)<\eta(j)$ or $\eta(i)\simeq\eta(j)$.
A pairing of the $E$ external lines is a set of $E/2$ pairs
\be
  \{ (e_1,e_{\frac{E}{2}+1}), \ldots, (e_{\frac{E}{2}},e_{E}) \},
\ee
with $e_1,\ldots,e_E \in \{ 1,\ldots ,E \}$ mutually distinct.
The set of all pairings is denoted by $\cP_\Gamma(E)$.

A path of lines of $\Gamma$ is a sequence
\be
  p= (w_1,l_1,w_2,l_2,\ldots,w_{n-1},l_{n-1},w_n)
\ee
with
\bea
   w_1,\ldots,w_n & \in & \cB_\Gamma \; ,\nonumber \\
   l_1,\ldots ,l_{n-1} & \in & \cL_\Gamma \; ,
    \mbox{mutually distinct}, \\
   w_i,w_{i+1} \in \widehat\Phi_\Gamma(l_i), &
   , & w_i\not= w_{i+1}\;\mbox{for all i=1,\dots,n-1}. \nonumber
\eea
If $w_1=w_n$, $p$ is called a loop of lines.
The set of loops of $\Gamma$ is denoted by $\cO_\Gamma$.
Two path of lines are called disjoint if they don't have any line
in common. $\cD_\Gamma$ is the set whose elements are the sets of
mutually disjoint paths.
Furthermore, the ordered tripel $(e_1,p,e_2)$ with
$e_1,e_2\in\{1,\ldots, E\}$, $e_1\not= e_2$ and
$\eta(e_1)=w_1,\eta(e_2)=w_n$, is called an external path of lines.
This does not exclude $p$ itself to be a loop. For,
$\eta(e_1)=\eta(e_2)$ is not excluded.
$\cE_\Gamma$ is the set of all external paths of $\Gamma$.
Finally, a set of paths $\{ p_1,\ldots, p_m \}$, $m\leq L$, is said
to cover $\Gamma$ if every internal line of $\Gamma$ belongs to
exactly one path.

With these notions, the O(N) symmetry number of $\Gamma$ is explicitly
given and determined by
\be
 C(\Gamma) =  \;
   \sum_{ P\in\cP_\Gamma(E)\;,\;
    {\{p_1,\ldots,p_{\frac{E}{2}}\}\in\cD_\Gamma} \atop
    (e_i,p_i,e_{\frac{E}{2}+i})\in\cE_\Gamma
    \;{\rm for\; all}\;(e_i,e_{\frac{E}{2}+i})\in P, i=1,\ldots,\frac{E}{2} }
  \sum_{0\leq M\leq L}
  \sum_{ q_1,\ldots ,q_M\in\cO_\Gamma \atop
    \{p_1,\ldots,p_{\frac{E}{2}},q_1,\ldots, q_M\} \;{\rm covers}\;\Gamma }
  N^M .
\ee
The actual recursive implementation of this formula from left to right is
straight for\-ward\footnote{This is one of the cases where the
description of an implementation is more complicated than the
implementation itself.}.
Paths and loops are constructed by moving systematically through the
graph.
Many simplifications can be taken into account for multiple lines
between vertices or multiple external lines at the same vertex.

\begin{figure}[htb]
\caption{\label{onreduction} Examples of graph reduction in the course
of the computation of the O(N) symmetry numbers. In the first case,
both $\mu$ and $\nu$ are odd numbers, and both are even in the
second example.}

\begin{center}
\setlength{\unitlength}{0.8cm}
\begin{picture}(10.0,6.0)


\put(0.0,3.0){
\setlength{\unitlength}{0.8cm}
\begin{picture}(4.6,3.0)

\put(0.0,2.0){\circle*{0.16}}
\put(0.0,2.0){\line(1,0){2.0}}
\put(0.0,2.0){\line(-1,1){0.5}}
\put(0.0,2.0){\line(-1,-1){0.5}}
\put(2.0,2.0){\circle*{0.16}}
\put(3.0,2.0){\oval(2.0,1.6)}
\put(4.0,2.0){\circle*{0.16}}
\put(4.0,2.0){\line(1,1){0.5}}
\put(4.0,2.0){\line(1,-1){0.5}}
\put(2.8,2.0){\makebox(0.4,0){$\mu$}}
\put(2.8,2.4){\makebox(0.4,0){$\cdot$}}
\put(2.8,1.6){\makebox(0.4,0){$\cdot$}}
\put(2.8,2.6){\makebox(0.4,0){$\cdot$}}
\put(2.8,1.4){\makebox(0.4,0){$\cdot$}}
\put(4.6,2.0){\makebox(0.4,0){$\nu$}}
\put(4.5,2.2){\makebox(0.4,0){$\cdot$}}
\put(4.5,1.8){\makebox(0.4,0){$\cdot$}}
\put(4.4,2.3){\makebox(0.4,0){$\cdot$}}
\put(4.4,1.7){\makebox(0.4,0){$\cdot$}}
\put(-0.75,2.0){\makebox(0.4,0){$\cdot$}}
\put(-0.7,2.2){\makebox(0.4,0){$\cdot$}}
\put(-0.7,1.8){\makebox(0.4,0){$\cdot$}}
\put(-0.6,2.3){\makebox(0.4,0){$\cdot$}}
\put(-0.6,1.7){\makebox(0.4,0){$\cdot$}}

\end{picture}
}


\put(6.3,3.0){
\setlength{\unitlength}{0.8cm}
\begin{picture}(1.0,3.0)

\put(0.0,2.0){\makebox(1.0,0){$\to\; f(N,\mu,\nu)\;$}}

\end{picture}
}


\put(9.3,3.0){
\setlength{\unitlength}{0.8cm}
\begin{picture}(3.0,3.0)

\put(0.0,2.0){\circle*{0.16}}
\put(0.0,2.0){\line(1,0){2.0}}
\put(0.0,2.0){\line(-1,1){0.5}}
\put(0.0,2.0){\line(-1,-1){0.5}}
\put(2.0,2.0){\circle*{0.16}}
\put(2.0,2.0){\line(1,1){0.5}}
\put(2.0,2.0){\line(1,-1){0.5}}
\put(2.6,2.0){\makebox(0.4,0){$\nu$}}
\put(2.5,2.2){\makebox(0.4,0){$\cdot$}}
\put(2.5,1.8){\makebox(0.4,0){$\cdot$}}
\put(2.4,2.3){\makebox(0.4,0){$\cdot$}}
\put(2.4,1.7){\makebox(0.4,0){$\cdot$}}
\put(-0.75,2.0){\makebox(0.4,0){$\cdot$}}
\put(-0.7,2.2){\makebox(0.4,0){$\cdot$}}
\put(-0.7,1.8){\makebox(0.4,0){$\cdot$}}
\put(-0.6,2.3){\makebox(0.4,0){$\cdot$}}
\put(-0.6,1.7){\makebox(0.4,0){$\cdot$}}

\end{picture}
}


\put(1.0,0.0){
\setlength{\unitlength}{0.8cm}
\begin{picture}(3.5,3.0)

\put(1.0,2.0){\circle*{0.16}}
\put(1.0,2.0){\line(-1,1){0.5}}
\put(1.0,2.0){\line(-1,-1){0.5}}
\put(3.0,2.0){\circle*{0.16}}
\put(2.0,2.0){\oval(2.0,1.6)}
\put(1.8,2.0){\makebox(0.4,0){$\mu$}}
\put(1.8,2.4){\makebox(0.4,0){$\cdot$}}
\put(1.8,1.6){\makebox(0.4,0){$\cdot$}}
\put(1.8,2.6){\makebox(0.4,0){$\cdot$}}
\put(1.8,1.4){\makebox(0.4,0){$\cdot$}}
\put(0.2,2.0){\makebox(0.4,0){$\nu$}}
\put(0.3,2.2){\makebox(0.4,0){$\cdot$}}
\put(0.3,1.8){\makebox(0.4,0){$\cdot$}}
\put(0.4,2.3){\makebox(0.4,0){$\cdot$}}
\put(0.4,1.7){\makebox(0.4,0){$\cdot$}}

\end{picture}
}


\put(6.0,0.0){
\setlength{\unitlength}{0.8cm}
\begin{picture}(1.0,3.0)

\put(0.0,2.0){\makebox(1.0,0){$\to\; g(N,\mu,\nu)\;$}}

\end{picture}
}


\put(8.0,0.0){
\setlength{\unitlength}{0.8cm}
\begin{picture}(3.0,3.0)

\put(1.0,2.0){\circle*{0.16}}
\put(1.0,2.0){\line(-1,1){0.5}}
\put(1.0,2.0){\line(-1,-1){0.5}}
\put(0.2,2.0){\makebox(0.4,0){$\nu$}}
\put(0.3,2.2){\makebox(0.4,0){$\cdot$}}
\put(0.3,1.8){\makebox(0.4,0){$\cdot$}}
\put(0.4,2.3){\makebox(0.4,0){$\cdot$}}
\put(0.4,1.7){\makebox(0.4,0){$\cdot$}}

\end{picture}
}

\end{picture}
\end{center}

\end{figure}
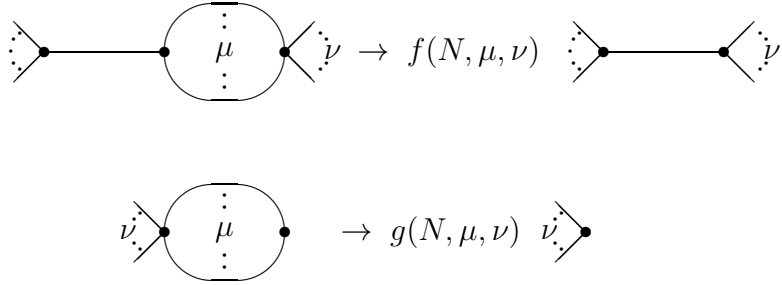


As is the case for the computation of the lattice imbedding numbers,
here also we don't have to compute the O(N) factor for every single
graph of a class. The first obvious observation to be made is that the
O(N) number does not depend on the length of 2-chains.
All the graphs that become topologically equivalent after shrinking
all 2-chains between non 2-vertices to a single line have identical
O(N) number. Other reductions are applied by explicitly
simplifying parts of the diagram that allow for closed computation.
Examples are given by Fig.~\ref{onreduction}.
Once a graph $\Gamma$ has been reduced to a graph
$\widetilde\Gamma$, its O(N) factor $C(\Gamma)$ is the product of

\begin{itemize}
\item the product of all $f$ and $g$ obtained during reduction,
denoted by $f_\Gamma(N)$, and
\item the O(N) factor $\widetilde C(\Gamma)$ of the reduced graph
$\widetilde\Gamma$.
\end{itemize}

Because many graph by reduction result into the same reduced graph
$\widetilde\Gamma$,
it is convenient to buffer $\widetilde\Gamma$ and $C(\widetilde\Gamma)$.
When a later graph $\Gamma^\prime$ of the class satisfies
$\widetilde{\Gamma^\prime}=\widetilde\Gamma$, then
\be
  C(\Gamma^\prime) = f_{\Gamma^\prime}(N) \cdot
  C(\widetilde\Gamma),
\ee
i.e.~$C(\widetilde\Gamma)$ does not need to be calculated a new.


\section{Results.}

We summarize the methods and the results obtained on susceptibility
series for the O(N) scalar field models
on the lattice, as described by the partition function
\eqn{lce.1}-\eqn{lce.4}.
The lattice is assumed to be of hypercubic geometry,
with periodic boundary conditions imposed for those directions in which
the extent is finite.
Fields at different lattice sites are coupled by nearest neighbour
pair interactions of strength $2\kappa$.


\begin{table}[htb]
\caption{\label{nrgph} The number of inequivalent graphs as yet generated
of the various graph classes
defined in the last sections. $L$ denotes the number of lines.
The renormalized moments $\cQ_{2n}$ have the same number of graphs
independent of the integer $n$. All the graphs are 1PI.
Both the renormalized moments and the 1VI graphs $\cS$ are even,
i.e.~every vertex has an even number of lines attached.
}
\vspace{0.5cm}

\begin{center}

\begin{tabular}{|r|r|r|r|r|r|r|}
\hline \hline
$L$ & $\cP_1(L)$ & $\cP_2(L)$ & $\cQ_2^{\rm ev}(L)$ &
$\cS_2^{\rm ev}(L)$ & $\cS_4^{\rm ev}(L)$ & $\cS_6^{\rm ev}(L)$ \\
[0.5ex] \hline

 0 & 1 & 1 & 1 & 1 & 1 & 1 \\
 1 & 0 & 0 & 0 & 0 & 0 & 0 \\
 2 & 1 & 1 & 1 & 0 & 1 & 1 \\
 3 & 0 & 1 & 0 & 1 & 1 & 2 \\
 4 & 2 & 3 & 4 & 0 & 4 & 6 \\
 5 & 0 & 3 & 0 & 2 & 4 & 11 \\
 6 & 5 & 11 & 15 & 3 & 20 & 46 \\
 7 & 1 & 16 &  0 & 8 & 27 & 91 \\
 8 & 15 & 53 & 79 & 9 & 117 & 349 \\
 9 & 7 & 112 &  0 & 40 & 214 & 837 \\
10 & 57 & 354 & 439 & 68 & 815 & 3140 \\ [0.5ex] \hline
11 & 48 & 953 &   0 & 247 & 1830 & 8401 \\
12 & 278 & 3160 & 2877 & 470 & 6721 & 31187 \\
13 & 379 & 9909 &    0 & 1779 & 17028 & 90599 \\
14 & 1647 & 34457 & 20507 & 3937 & 61653 & 336582 \\ [0.5ex] \hline
15 & 3328 & 119921 &    0 & 14801 & 170923 & 1042392 \\
16 & 12321 & 439552 & 161459 & 35509 & 621191 & 3895341 \\
17 & 31869 & 1638878 &     0 & 135988 & 1834324 &  \\
18 & 111493 & 6312209 & 1376794 & 350614 & 5548427 &  \\
19 & 337817 & 24810845 &     0 & 1361878 & 20967387 &  \\ \hline
\end{tabular}

\end{center}

\end{table}


Susceptibilities are given as power series with respect to the
coupling strength $2\kappa$, with coefficients depending on the bare
coupling constants $\lambda$.
These series have a non-vanishing radius
of convergence.
The coefficients are represented as sums over equivalence classes of
connected graphs with appropriate weights. The number of internal lines of
a diagram is equal to the order of $2\kappa$ it contributes to.

With increasing order, the construction and computation of all the
connected graphs of a correlation function or
susceptibility becomes managable only if strict bookkepping
is done.
Both the number of graphs and their complexity are considerably reduced by
applying partial resummations. The first essential step in this
direction is the representation of connected correlations in terms of
one-particle irreducible (1PI) ones. The latter are sums of 1PI graphs
only and have similar series representations,
\be \label{res.1}
  \chi_E^{\rm 1PI} = \sum_{L\geq 0} a_{E,L}^{\rm 1PI}(\lambda) \;
  (2\kappa)^L,
\ee
and similar for weighted correlations such as
\be \label{res.1.1}
  \mu_2^{\rm 1PI} = \sum_{L\geq 0} \mu_{2,L}^{\rm 1PI}(\lambda) \;
   (2\kappa)^L,
\ee
cf.~\eqn{lce.10.2},\eqn{lce.28.11}.
For instance, 2-point susceptibilities are related by
\be \label{res.2}
  \chi_2 = \frac{\chi_2^{\rm 1PI}}{1-(2\kappa)(2D)\chi_2^{\rm 1PI}} ,
\ee
cf.~(\ref{lce.28.10}f).
A further useful reorganization is some kind of product representation
of weights in terms of graphs that are in addition one-vertex
irreducible, and so-called renormalized moments,
the classes $\cS$ and $\cQ$. It is sufficient to construct the latter
ones explicitly. The expensive computation of the weights of graphs
must be done for those classes only.

In order to manage diagrams numerically, every
equivalence class of graphs is represented by an incidence matrix.
This representation is made unique by defining a total order relation
on the set of symmetric matrices, and than invoke an appropriate
maximization procedure over all simultaneous permutations
of rows and columns. Those permutations correspond to different
enumeration of vertices. To avoid huge factorials of permutations to be
carried out, first of all order relations on vertices are introduced,
leading to a preordered incidence matrix. The canonical incidence matrix
then is defined by maximizing only over those permutations which
exchange vertices that are left unordered relative to each
other.

It is at the heart of the construction mechanism
to make this representation as efficient as possible.
We have to introduce an extraordinary far-going ordering on vertices
to obtain all the graphs
contributing to the susceptibility series
up to 18th order and beyond.
The numbers of the graphs as yet generated are listed in
Tab.~\ref{nrgph}. For the O(N) scalar models, only those graphs
have non-vanishing weights that have even number of lines at every vertex.
In turn the number of external lines is even as well.


\begin{table}[htb]
\caption{\label{vertstruct} Examples for the number of vertex structures
of various graph classes.
}

\begin{center}

\begin{tabular}{|r|r|r|r|r|r|}\hline\hline
$L$ & $\cQ_2^{\rm ev}(L)$ & $\cQ_4^{\rm ev}(L)$ &
$\cS_2^{\rm ev}(L)$ & $\cS_4^{\rm ev}(L)$ & $\cS_6^{\rm ev}(L)$ \\
[0.5ex] \hline

 0 &   1 &   1 &   1 &   1 &   1  \\
 1 &   0 &   0 &   0 &   0 &   0  \\
 2 &   1 &   1 &   0 &   1 &   1  \\
 3 &   0 &   0 &   1 &   1 &   2  \\
 4 &   4 &   4 &   0 &   3 &   4  \\
 5 &   0 &   0 &   2 &   4 &   8  \\
 6 &  10 &  11 &   3 &   9 &  14  \\
 7 &   0 &   0 &   6 &  10 &  20  \\
 8 &  21 &  27 &   7 &  19 &  29  \\
 9 &   0 &   0 &  11 &  24 &  42  \\
10 &  42 &  50 &  19 &  39 &  60  \\ [0.5ex] \hline
11 &   0 &   0 &  28 &  48 &  81  \\
12 &  78 & 102 &  39 &  74 & 110  \\
13 &   0 &   0 &  49 &  92 & 147  \\
14 & 139 & 173 &  78 & 135 & 198  \\ [0.5ex] \hline
15 &   0 &   0 & 102 & 169 & 259  \\
16 & 242 & 312 & 146 & 238 & 338  \\
17 &   0 &   0 & 178 & 297 &  \\
18 & 408 & 505 & 264 & 404 &  \\ \hline
\end{tabular}

\end{center}

\end{table}


Storage of that many graphs or incidence matrices
is as binary words, using a kind of Huffmann tree coding mechanism.
It is done in such a way that all graphs with same number of internal,
external lines and vertices have identical bit lengths. This is most
convenient for the many cases graphs have to be compared. What is
compared actually are their binary words.

The computation of the weights of the graphs is done as outlined in
the last section.
The laborious part here consists of the determination of lattice
imbedding numbers and O(N) symmetry factors.
In total, two orders increase the number of graphs of the classes considered
by a factor of about 10. Increasing complexity implies a total cost
factor of about 30-40 to compute two orders more.

Two final problems have to be managed.
Vertices contribute to the weights in dependence on the bare
coupling constants. In principle, summation of millions of graphs must
be done each time a coupling constant changes. Even worse, the weights
typically are of comparable size, so we have to be aware of severe
roundoff errors.
The total weight of a graph consists of two factors. The first is a
non-negative rational number, itself a product of lattice imbedding
number and O(N) and topological symmetry numbers.
The second one comes from the product over the vertices.
Every vertex contributes a finite
dimensional integral that in most cases can be done numerically only.
These integrals alternate in sign with respect to half the number of lines
entering the vertex.

We solve the problem of both having to do a huge sum each time the
couplings change and the accumulation of roundoff errors by introducing
so-called vertex structures.
A vertex structure maps a graph onto a multitupel of non-negative
integers,
\bea \label{res.3}
  \cV & : & \cG \to ( \{ 0,1,2,\ldots\} )^n , \\
  & & \Gamma \to (\mu_1,\mu_2,\ldots ). \nonumber
\eea
Here, $n$ is an upper bound on the number of the sum of internal
and external lines of a graph under consideration, and
$\mu_i$ denotes the number of vertices of the graph $\Gamma$ that have
precisely $i$ lines entering, i.e. of
\be \label{res.4}
  \{ v\in\cB_\Gamma \; \vert \; l(v)=i \} .
\ee
The multitupel $(\mu_1,\mu_2,\ldots )$ is called the vertex structure
of $\Gamma$.
For the O(N) case, only $\mu_i$ with even $i$ can be different from
zero.


\begin{table}[htb]
\caption{\label{sus4d4} Coefficients of susceptibility series
for the O(4) model at $\lambda_1=\infty$ and
$\lambda_2=0$, on the $4\times\infty^3$ lattice.
}
\vspace{0.5cm}

\begin{center}

\begin{tabular}{|r|r|r|r|r|}\hline \hline
$L$ & $a_{2,L}^{\rm 1PI}$ & $\mu_{2,L}^{\rm 1PI}$ &
$a_{4,L}^{\rm 1PI}$ & $a_{6,L}^{\rm 1PI}$ \\[0.5ex] \hline

 0 & 0.25          & 0            &-0.0625      &  0.078125     \\
 1 & 0             & 0            & 0           &  0            \\
 2 & -0.125        & 0            & 0.1875       & -0.6640625000\\
 3 & 0.01041666667 &0.0078125     &-0.04166666667& 0.2669270833\\
 4 & 0.01627604167 &0             &-0.1513671875 & 1.236368815 \\
 5 & 0.001139322917&0.0004882812500&0.01839192708&-0.3427937826\\
 6 &-0.03774685330 &0.002115885417& 0.1999131944 &-2.336894565 \\
 7 & 0.008895534939&0.006170654297&-0.08401557075& 1.321710798 \\
 8 &-0.03999481201 &0.004740397135& 0.04620615641& 0.8677455584\\
 9 & 0.01268447593 &0.009544584486&-0.08312718427& 0.5234833748\\
10 &-0.08279821256 &0.01060011122&  0.3628878114& -5.029520772\\ [0.5ex] \hline
11 & 0.02792425547 &0.02405914339& -0.3008789073&  5.515860626\\
12 &-0.1571703581 & 0.02491494020&  0.7014666549& -7.989323468\\
13 & 0.05760043398 &0.05506860056& -0.6772442574& 11.90025208 \\
14 &-0.3147585986 & 0.06230568066&  1.703826430 &-25.49575660 \\ [0.5ex] \hline
15 & 0.1235842297 & 0.1303351125 & -1.691767724 & 36.94182961 \\
16 &-0.6430290985 & 0.1554514202 &  3.999895232 &-70.54713230 \\
17 & 0.2684262835 & 0.3123609133 & -4.227999178 &  \\
18 &-1.343451682 &  0.3887314574 &  9.417754413 &  \\ \hline
\end{tabular}

\end{center}

\end{table}



\begin{table}[htb]
\caption{\label{sus4d6} As in Tab.~3, but on the
$6\times\infty^3$ lattice.
}
\vspace{0.5cm}

\begin{center}

\begin{tabular}{|r|r|r|r|r|}\hline \hline
$L$ & $a_{2,L}^{\rm 1PI}$ & $\mu_{2,L}^{\rm 1PI}$ &
$a_{4,L}^{\rm 1PI}$ & $a_{6,L}^{\rm 1PI}$ \\[0.5ex] \hline

 0 & 0.25           & 0              &-0.0625        & 0.078125     \\
 1 & 0              & 0              & 0             & 0            \\
 2 & -0.125         & 0              & 0.1875        &-0.6640625000 \\
 3 & 0.01041666667  & 0.0078125      &-0.04166666667 & 0.2669270833 \\
 4 & 0.01822916667  & 0              &-0.1562500000  & 1.263834635  \\
 5 & 0.0006510416667& 0.0004882812500& 0.02148437500 &-0.3728027344 \\
 6 &-0.03359646267  & 0.002115885417 & 0.1898695204  &-2.278199090  \\
 7 & 0.007688395182 & 0.005743408203 &-0.07582329644 & 1.235062069  \\
 8 &-0.02836252848  & 0.004374186198 & 0.004855290166& 1.245049212  \\
 9 & 0.009167494597 & 0.007763120863 &-0.05144329777 & 0.06869500487\\
10 &-0.05681311643  & 0.008740446303 & 0.2625076545  &-4.044370528  \\ [0.5ex]
\hline
11 & 0.01901101339  & 0.01786831712  &-0.2090660888  & 4.051910630  \\
12 &-0.1080852959   & 0.01844869455  & 0.4873771492  &-5.358685469  \\
13 & 0.03790109259  & 0.03792749792  &-0.4478781738  & 7.732701464  \\
14 &-0.2258664596   & 0.04196657999  & 1.260000420   &-18.95164927  \\ [0.5ex]
\hline
15 & 0.08196611346  & 0.08673188894  &-1.145127753   & 25.74367205  \\
16 &-0.4902317790   & 0.09916992312  & 3.140756286   &-55.14063511  \\
17 & 0.1833679347   & 0.2045461221   &-2.979792099   &  \\
18 &-1.094178613    & 0.2415480079   & 7.829428325   &  \\ \hline
\end{tabular}

\end{center}

\end{table}



\begin{table}[htb]
\caption{\label{sus3d} As in Tab.~3, but on the 3-dimensional
$\infty^3$ lattice.
}
\vspace{0.5cm}

\begin{center}

\begin{tabular}{|r|r|r|r|r|}\hline \hline
$L$ & $a_{2,L}^{\rm 1PI}$ & $\mu_{2,L}^{\rm 1PI}$ &
$a_{4,L}^{\rm 1PI}$ & $a_{6,L}^{\rm 1PI}$ \\[0.5ex] \hline

 0 & 0.25           & 0              &-0.0625        &  0.078125     \\
 1 & 0              & 0              & 0             &  0            \\
 2 &-0.09375        & 0              & 0.140625      &-0.4980468750  \\
 3 & 0.0078125      & 0.0078125      &-0.03125       & 0.2001953125  \\
 4 & 0.01367187500  & 0              &-0.09472656250 & 0.7391357422  \\
 5 &-0.0004882812500&-0.0004882812500& 0.01708984375 &-0.2545776367  \\
 6 &-0.008585611979 & 0.001464843750 & 0.06603190104 &-0.8673782349  \\
 7 & 0.001572672526 & 0.001572672526 &-0.02043457031 & 0.3948656718  \\
 8 & 0.0003199259440& 0.001302083333 &-0.02937910292 & 0.6583463351  \\
 9 & 0.0001472897000& 0.0005287594265& 0.007233217027&-0.3005849621  \\
10 &-0.001710902320 & 0.0008281707764& 0.02126218503 &-0.5705569410  \\ [0.5ex]
\hline
11 & 0.0005031175714& 0.0009918579980&-0.01068644069 & 0.3722756644  \\
12 &-0.0007259439854& 0.0008832578306&-0.002698613418& 0.2289861101  \\
13 & 0.0002879108938& 0.0008595423960&-0.001406942080&-0.1410064280  \\
14 &-0.0007758281759& 0.0008468480700& 0.007465842599&-0.1955813214  \\ [0.5ex]
\hline
15 & 0.0002983264770& 0.0008996308209&-0.004516607350& 0.2479154031  \\
16 &-0.0005740168913& 0.0008895557789& 0.005824725792&-0.3629113447  \\
17 & 0.0002457315931& 0.0008854585355&-0.01304684165 &  \\
18 &-0.0006825215700& 0.0008996335601& 0.003926643743&  \\ \hline
\end{tabular}

\end{center}

\end{table}


The advantage of introducing this notion is that the number of
different vertex structures is much smaller than the number of
topologically inequivalent graphs.
Except for factors contributed by the vertices themselves,
we sum the (positive) weights of all graphs
with the same vertex structure. This yields weights
that do no more depend on the bare coupling constants.
For instance, in this way, the coefficients of \eqn{res.1} become
\be \label{res.5}
  a_{E,L}^{\rm 1PI}(\lambda) = \sum_{ (\mu_1,\mu_2,\ldots ) }
   b_{E,L}^{\rm 1PI}(\mu_1,\mu_2,\ldots ) \cdot
   \prod_{i\geq 1} \left(
     \stackrel{\circ}{v}_{i}^c(\lambda) \right)^{\mu_i},
\ee
where the $b_{E,L}^{\rm 1PI}$ do not depend on the coupling constants.

Some numbers of vertex structures are listed in Tab.~\ref{vertstruct}.
For a given lattice, including dimension and topology, and given
symmetry group O(N), we don't need to keep all the graphs.
Only those vertex structures together with
the small set of numbers $b_{E,L}^{\rm 1PI}(\mu_1,\mu_2,\ldots )$
are needed for the final evaluation of the LCE series coefficients
such as \eqn{res.5}, for all values of the coupling constants $\lambda$.

As an example of the resulting series we consider the O(4)-symmetric
$\Phi^4$-theory, as described by the action \eqn{lce.1}-\eqn{lce.4},
with $\lambda_1=\infty$ and $\lambda_2=0$.
The coefficients of the various 1PI susceptibilities are listed in the
tables Tab.~\ref{sus4d4}, \ref{sus4d6} and \ref{sus3d} for the
hypercubic lattices of the sizes
$4\times\infty^3$, $6\times\infty^3$ and $\infty^3$, respectively.

It appears that
the 1PI series are often of better use than the connected
ones. For instance, the critical line $\kappa_c(\lambda_1)$
can be determined as the smallest positive root of
\be \label{res.6}
  \chi_2(\lambda_1)^{-1} =
    \frac{1-(2\kappa)(2D)\chi_2^{\rm 1PI}(\lambda_1)}
     {\chi_2^{\rm 1PI}(\lambda_1)}
\ee
with significantly better accuracy than from the various ratio and
Pade' methods applied to the series of $\chi_2(\lambda_1)$.

The four-dimensional finite temperature data yield the following
numbers on the standard high temperature critical exponents:
\pagebreak
\bea
   \gamma & = & 1.443(25) , \nonumber \\
   \nu & = & 0.735(11) , \\
   \eta & = & 0.035(28), \nonumber
\eea
whereas the three-dimensional series yield
\bea
   \gamma & = & 1.449(16) , \nonumber \\
   \nu & = & 0.735(8) , \\
   \eta & = & 0.029(17). \nonumber
\eea
Derivation and discussion of the above exponents as well as other
critical data
and the presentation of the methods used will be given elsewhere.
Our aim here was to present those techniques that are capable of
pushing the linked cluster expansion series towards very high orders,
in such a way that details on critical behaviour can be obtained,
even on asymmetric (finite temperature) lattices.

\section*{Acknowledgement}

Part of this work was done during my stay at CERN as Scientific Associate.
I am indebted to A.~Pordt for providing the convergence proof
of LCE series prior to publication.


\section{Extended vertex ordering for non-exact ring graphs.}

In this appendix we list the notions needed to define the order relation
$\cO$ of \eqn{alg.8} for the vertices of a non-exact ring graph,
as announced in subsection \ref{inc.ring}. Essentially, the notion
of an internal 2-vertex is to be replaced by symmetric 2-vertex.

Hence let $\Gamma$ be a non-exact ring graph. We define
\be \label{exring.1}
   \cO = < \cO_{LW}, [\cO_{ns},\cO_{symm}] >.
\ee
$\cO_{symm}$ is an order relation on $\cB_{\Gamma,symm}^{(2)}$, the set
of symmetric 2-vertices of $\Gamma$, and$\cO_{ns}$ an order relation on
$\cB_\Gamma\setminus\cB_{\Gamma,symm}^{(2)}$.
They are defined along the following lines.

A symmetric 2-chain of $v\in\cB_{\Gamma,symm}^{(2)}$ is a path
$p=(w_1,w_2,\ldots,w_n)$ with $w_1=v$ and
$w_2,\ldots,w_{n-1}\in\cB_{\Gamma,symm}^{(2)}$ mutually distinct
vertices. If the last vertex is
not a symmetric 2-vertex, $w_n\not\in\cB_{\Gamma,symm}^{(2)}$, $p$ is
called maximal symmetric 2-chain of $v$.
If $v\in\cB_{\Gamma,symm}^{(2)}$,
we assign a tripel of two integers and a vertex to p by
\bea
  \mu(p) & = & n-1 , \\ \label{alg.14}
  \lambda(p) & = & c(v) ,\quad\mbox{and} \\
  e(p) & = & w_n.
\eea
Every $v\in\cB_{\Gamma,symm}^{(2)}$ has precisely two distinct
maximal symmetric 2-chains. We denote them by $c_1(v)$ and $c_2(v)$.
They intersect in $v$ and possibly the last vertex only.
Furthermore we set
\be \label{exring.2}
  \mu_{symm}(v) = \max_{i=1,2} \; \mu(c_i).
\ee
Let $\cO_{non}^\prime$ be the order relation on
$\cB_\Gamma\setminus\cB_{\Gamma,symm}^{(2)}$, defined by
$\cO_{non}^\prime = < \cO_\delta, \cO_{ns} >$,
where
\be
  v\geo{\cO_\delta}w \Longleftrightarrow
   \max_{\nu_k(v)\not=0} k  > \max_{\nu_k(w)\not=0} k,
\ee
for all $v,w\in\cB_\Gamma\setminus\cB_{\Gamma,symm}^{(2)}$.
It induces an order on the
maximal symmetric 2-chains $\{c_1(v),c_2(v)\}$ by
\be \label{exring.3}
  c_i(v) > c_j(v) \; \Longleftrightarrow \;
   \begin{array}{ c }
    \mu(c_i) < \mu(c_j) \; \mbox{, or} \\
    \mu(c_i) = \mu(c_j) \; \mbox{and} \; \lambda(c_i)>\lambda(c_j)
     \; \mbox{, or} \\
    \mu(c_i) = \mu(c_j) \; \mbox{and} \; \lambda(c_i)=\lambda(c_j)
     \; \mbox{and}\; e(c_i)\geo{\cO_{non}^\prime}e(c_j) .
   \end{array}
\ee
If unique, we choose "short and long chains" by
\bea
  c_s(v) & = & \max_{i=1,2} \; c_i(v), \nonumber \\ \label{exring.4}
  c_l(v) & = & \min_{i=1,2} \; c_i(v).
\eea
If neither $c_1(v)>c_2(v)$ nor $c_2(v)>c_1(v)$, we choose any one of the
maximal symmetric 2-chains as $c_s(v)$ and the other one as $c_l(v)$.
Again, this freedom is allowed because all that will be used is the order
relation \eqn{exring.3}.

$\cO_{non}^\prime$ induces through \eqn{exring.4} an order relation
$\cO_{symm}$ on $\cB_{\Gamma,symm}^{(2)}$ by
$\cO_{symm} = < \cO_{i1}, \cO_{i2} >$, with
\bea \label{exring.5}
  v\geo{\cO_{i1}}w & \Longleftrightarrow &
   c_s(v) > c_s(w) , \\
  v\geo{\cO_{i2}}w & \Longleftrightarrow &
   c_l(v) > c_l(w) ,
\eea
for all symmetric 2-vertices $v,w$.
Finally, $\cO_{ns}$ is defined as follows. With
\bea
  s_1(v) & = & \sum_{w\in\cN(v)\atop
    w\in\cB_{\Gamma,symm}^{(2)}} \mu_{symm}(w) , \\
  s_2(v) & = & \max_{w\in\cN(v)\atop
    w\in\cB_{\Gamma,symm}^{(2)}} \mu_{symm}(w) ,
\eea
for every $v\not\in\cB_{\Gamma,symm}^{(2)}$, we set
$\cO_{ns} = < \cO_{ns,1},\cO_{ns,2} >$, where
\be
  v \geo{\cO_{ns,k}}w \; \Longleftrightarrow \;
  s_k(v) > s_k(w),
\ee
for $k=1,2$ and
for all $v,w\in\cB_\Gamma\setminus\cB_{\Gamma,symm}^{(2)}$.


%
%


\end{document}